\documentclass{aa}  

\usepackage{graphicx}
\usepackage{txfonts}
\usepackage{natbib}
\bibpunct{(}{)}{;}{a}{}{,} % to follow the A&A style
\usepackage{booktabs}
\usepackage{multicol}
\usepackage{graphicx}
\usepackage{fancyhdr}
\usepackage[hidelinks,colorlinks=true,linkcolor=blue,citecolor=blue]{hyperref}
\usepackage{amsmath}	% Advanced maths commands
\usepackage{amssymb}	% Extra maths symbols
\usepackage[inter-unit-product=\cdot]{siunitx}
\usepackage{enumerate}
\usepackage{multirow}
\usepackage{mathtools}
\usepackage{longtable}
\usepackage{tabularx}
\usepackage{xcolor} % Colors used for comments commands
\usepackage{ulem} % Cancel text with horizontal bar
\usepackage{comment}
\usepackage{soul}

\newcommand{\Msun}{\,\mathrm{M}_\odot}

\newcommand{\MSMBH}{M_\bullet}

\newcommand{\fEdd}{\lambda_\bullet}

\newcommand{\quadre}[1]{\left[#1\right]}

\newcommand{\fastcluster}[1]{{\fontfamily{lmtt}\selectfont fastcluster}}
\newcommand{\fastclusterAGN}[1]{{\fontfamily{lmtt}\selectfont fastclusterAGN}}
\newcommand{\sevn}[1]{{\fontfamily{lmtt}\selectfont sevn}}
\newcommand{\pagn}[1]{{\fontfamily{lmtt}\selectfont pAGN}}
\newcommand{\tsunami}[1]{{\fontfamily{lmtt}\selectfont tsunami}}
\newcommand{\mcfacts}[1]{{\fontfamily{lmtt}\selectfont McFacts}}
\newcommand{\CAT}[1]{{\fontfamily{lmtt}\selectfont cat}}
\newcommand{\AGNrates}[1]{{\fontfamily{lmtt}\selectfont AGN$\mathcal{R}$ates}}

\newcommand{\orcidicon}[1]{\href{https://orcid.org/#1}{\includegraphics[width=11pt]{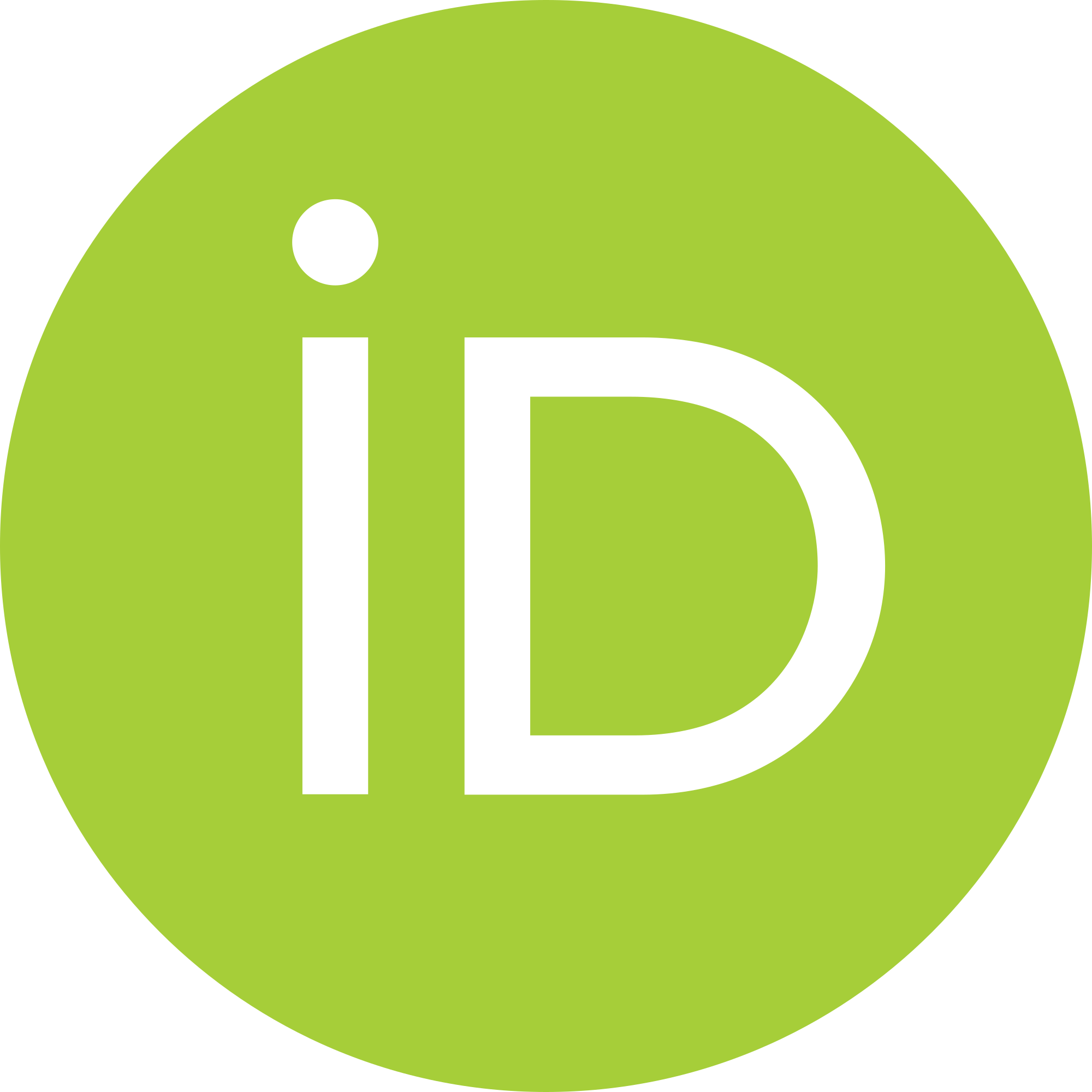}}}
\newcommand{\orcid}[1]{\href{https://orcid.org/#1}{\protect\orcidicon{#1}}}

\DeclareSIUnit\year{yr}
\DeclareSIUnit\au{AU}
\DeclareSIUnit\parsec{pc}
\DeclareSIUnit\erg{erg}

\defcitealias{SG}{SG}
\defcitealias{TQM}{TQM}
\defcitealias{Bellovary_2016}{B16}
\defcitealias{Grishin_2024}{G24}
\defcitealias{Ishibashi_2020}{IG20}
\defcitealias{Ishibashi_2024}{IG24}
\defcitealias{Calcino_2024}{C24}

\begin{document} 

%%%%%%%%%%%%%%%%%%%%%%%%%%%%%%%%%%%%%%%%
% if you use custom commands in your title,
% ensure to check your title when submitting!
%%%%%%%%%%%%%%%%%%%%%%%%%%%%%%%%%%%%%%%%
   \title{AGN-driven BBH mergers: Black hole populations \\ and hierarchical growth across the AGN parameter space}
   \titlerunning{AGN-driven BBH mergers}

%%%%%%%%%%%%%%%%%%%%%%%%%%%%%%%%%%%%%%%%
% Please separate each author with the \and command
%
% Please do not include ORCIDs next to author names.
% Only ORCIDs authenticated by individual authors in EDPS
% editorial system will be taken into account.
% ORCIDs included here will be removed.
%%%%%%%%%%%%%%%%%%%%%%%%%%%%%%%%%%%%%%%%

   \author{M. Paola Vaccaro\inst{1,2}\orcid{0000-0003-3776-9246} \and Michela Mapelli\inst{1,2,3,4}\orcid{0000-0001-8799-2548} \and Alessandro A. Trani\inst{5,6}\orcid{0000-0001-5371-3432} \and Boyuan Liu\inst{1,2}\orcid{0000-0002-4966-7450}
        }
    \authorrunning{M.P. Vaccaro et al.}
   \institute{Institut f{\"u}r Theoretische Astrophysik, ZAH, Universit{\"a}t Heidelberg, Albert-Ueberle-Stra{\ss}e 2, D-69120, Heidelberg, Germany 
    \and
    Interdisziplin{\"a}res Zentrum f{\"u}r Wissenschaftliches Rechnen, Universit{\"a}t Heidelberg, Heidelberg, Germany
    \and
    Physics and Astronomy Department Galileo Galilei, University of Padova, Vicolo dell'Osservatorio 3, I-35122, Padova, Italy
    \and
    INFN, Sezione di Padova, Via Marzolo 8, I-35131, Padova, Italy
    \and 
    Department of Astronomy, University of Concepción, Avenida Esteban Iturra s/n Casilla 160-C Concepción, Chile 
    \and 
    INFN, Sezione di Trieste, I-34127, Trieste, Italy\\ 
    \email{\href{mailto:mariapaolavaccaro@gmail.com}{mariapaolavaccaro@gmail.com}, \href{mailto:mapelli@uni-heidelberg.de}{mapelli@uni-heidelberg.de}}
    }

   \date{Received XXX; accepted YYY}

% \abstract{}{}{}{}{}
% 5 {} token are mandatory
 
  \abstract
   {
Active galactic nuclei (AGNs) have been proposed as efficient environments for the formation of binary black holes (BBHs).
We present an updated semi-analytical framework for BBH formation and evolution in AGN disks, following the capture, migration, pair-up, gas-driven hardening, binary--single encounters, and merger of stellar-origin black holes. 
We systematically explore the dependence of the resulting BBH merger population on the main AGN parameters, namely the supermassive black hole mass $\MSMBH$, the Eddington ratio $\fEdd$, and the disk viscosity parameter $\alpha$, and construct an intrinsic BBH population by weighting the simulations according to observed low-redshift AGN properties. 
We find that AGN disks can produce repeated mergers and build a high-mass tail extending beyond the pair-instability mass gap and into the intermediate-mass range. 
Hierarchical growth is more efficient in lower-viscosity disks, with $\alpha=0.01$, while higher-viscosity disks suppress the formation of massive remnants. 
The merger efficiency generally increases with $\fEdd$, but its dependence on $\MSMBH$ is non-trivial. 
The AGN-assisted BBH population is characterized by increasingly unequal mass ratios at high primary mass, a correlation between primary mass and $|\chi_{\rm eff}|$, and an effective-spin distribution that depends strongly on the fraction of binaries born in prograde or retrograde configurations. 
Most mergers enter the ground-based detector %band
frequency range with low eccentricities, $e \sim 10^{-3}$--$10^{-2}$ at $f_{\rm peak}=10\,{\rm Hz}$, although a subpopulation retains high eccentricity.
We find that the AGN channel can reproduce systems broadly consistent with the massive BBH events GW190521 and GW231123. 
We test several variations of the physical model, including different formalisms for migration torques, gas hardening, and three-body encounters. 
The general properties of the population are robust across these variations, with the high-mass tail and spin signatures persisting in all cases except when gas hardening is switched off. This confirms that gas-driven binary evolution is a key ingredient for efficient hierarchical growth in AGN disks.
   }


   \maketitle
   \nolinenumbers

%%%%%%%%%%%%%%%%%%%%%%%%%%%%%%%%%%%%%%%%%%%%%%%%%%%%%%%%%%%%%%
\section{Introduction}

Gravitational-wave (GW) astronomy has revealed a large and growing population of binary black hole (BBH) mergers \citep{GWTC5_pop, GWTC5_catalog, GWTC4_pop, GWTC4_catalog}. These observations have established BBH mergers as a powerful laboratory to test general relativity and probe compact object astrophysics, %as a common astrophysical phenomenon, 
while simultaneously revealing a broad diversity in their masses, spins, and orbital properties. A variety of formation channels have been proposed to explain this population \citep{Mapelli2021_review, Mandel_2022, Tauris_2023}, including isolated binary evolution \citep{Dominik_2012, Belczynski_2016, sevn_2017, Tanikawa_2022b, sevn_2023} and dynamical environments such as stellar clusters \citep{Portegies_Zwart_2000,Mapelli_2016,Fragione_Kocsis_2018, Kumamoto_2019, Rastello_2019, Atallah_2023, Chattopadhyay_2023, ArcaSedda_2026} and active galactic nuclei (AGN) disks \citep{McKernan_2012,Bartos_2017, Stone_2017, Yang_2019, Secunda_2020, Tagawa_2020_b, Tagawa_2026c, Samsing_2022, Vaccaro_2023, McFacts3_2024, Dittmann_2025, Xue_2025}.

Among the proposed dynamical channels, AGN disks have emerged as a promising environment for BBH formation and mergers. Stellar-mass black holes (BHs) embedded in AGN disks can migrate through the gaseous medium, pair up, and subsequently harden and merge under the combined action of gas torques, few-body encounters, and GW emission. The AGN channel naturally predicts distinctive merger properties, including high masses, hierarchical mergers, correlations between masses and spins, and potentially non-negligible eccentricities in the sensitivity band of ground-based GW detectors \citep{Samsing_2022, Santini_2023, Vaccaro_2023, McFacts2_2024}. In addition, mergers occurring close to the central supermassive black hole (SMBH) may be affected by strong lensing, Doppler shifts, or peculiar accelerations \citep{Tamanini_2020, Leong_2025, Samsing_2025, Tagawa_2026b} and may be accompanied by electromagnetic flares \citep{Graham_gw190521, Morton_2023, Joshi_2025, Tagawa_2023, Tagawa_2026, McFacts4_2026}, further enriching the phenomenology of this channel.

Recent observational studies have shown that the AGN channel cannot account for the entirety of the BBH merger population reported by current GW observations,
%a non-negligible fraction of the BBH merger population reported by current GW observations may be compatible with an AGN origin, 
with inferred upper limits at the level of $\sim 10 \%$ \citep{Fishbach_2022, Veronesi_2023, Veronesi_2025b, Veronesi_2025, Vaccaro_2023, Afroz_2025,
Cabrera_2026, Padhyegurjar_2026}. At the same time, the predicted properties of AGN-assisted BBH mergers remain highly uncertain. Recent hydrodynamical simulations and semi-analytical studies have highlighted the strong dependence of this channel on uncertain physical ingredients, including the AGN disk structure, migration torques, gas hardening efficiency, AGN lifetime, and the abundance and spatial distribution of embedded BHs \citep[e.g.,][]{Tagawa_2020, Tagawa_2021, Tagawa_2023, Whitehead_2025b, Rowan_2023, Rowan_2025b, McFacts1_2024, McFacts3_2024, Trani_2024, Vaccaro_2023, Vaccaro_2025}. Different studies often adopt different assumptions for these ingredients, making their predictions difficult to compare directly. A systematic exploration of the AGN parameter space is therefore essential to assess the robustness of BBH merger predictions, identify the dominant sources of uncertainty, and determine the most distinctive observational signatures of this channel.

Here, we construct a comprehensive and up-to-date semi-analytical framework to model the formation, evolution, and merger of BBHs in AGN disks, publicly available within the \fastcluster{} software environment\footnote{\fastcluster{} is an open-source code hosted on GitLab at
\href{https://gitlab.com/micmap/fastcluster_open}{\texttt{fastcluster\_open}}.
The \fastclusterAGN{} version used in this work is available at
\href{https://gitlab.com/mariapaola.vaccaro/fastcluster_agn}{\texttt{fastcluster\_agn}}.} \citep{fastcluster2021, fastcluster2022, Vaccaro_2023, Torniamenti_2024}. Starting from populations of stellar-mass BHs, we follow their capture, migration, pair-up, hardening, and merger in AGN disks, accounting for both gas-driven and dynamical interactions. Relative to our previous work \citep{Vaccaro_2023}, we incorporate updated prescriptions motivated by recent hydrodynamical and few-body studies, and systematically explore how different assumptions on AGN disk structure, migration physics, gas hardening, SMBH mass, and AGN activity cycles affect the resulting BBH merger population. %We follow repeated mergers and hierarchical growth, and investigate not only the effect of individual parameters, but also their interplay in shaping the observable BBH population.

Our analysis shows that the AGN channel does not lead to a single characteristic BBH population, but instead spans a broad range of outcomes across the AGN parameter space. In particular, hierarchical growth is most efficient in low-viscosity disks and can produce a high-mass tail extending beyond the pair-instability mass gap, while high-viscosity disks strongly suppress the formation of the most massive remnants. We also identify distinctive population-level signatures of AGN-assisted mergers, including increasingly unequal mass ratios at high primary mass, a correlation between primary mass and $|\chi_{\rm eff}|$, and an effective-spin distribution that retains memory of the initial prograde or retrograde orbital configuration. These features persist across several model variations, except when gas hardening is removed, highlighting the central role of gas-driven binary evolution in enabling hierarchical growth in AGN disks.

%The next generation of ground-based GW detectors, including the Einstein Telescope (ET) and Cosmic Explorer (CE) \citep{Punturo_2010_ET, Maggiore_2020_ET, evans2021_CE, Branchesi_2023_ET}, will allow us to detect BBH mergers out to redshifts of $z \sim 100$ \citep{Hall_2019}. Interpreting these future observations will require models that consistently connect BBH formation channels to the evolution of AGNs and SMBHs across cosmic time. This is particularly important because the growth of SMBHs in the early Universe remains highly uncertain \citep{Volonteri_2010}, especially after recent JWST observations of Little Red Dots at high redshift that may host SMBHs with masses above $10^6\Msun$ at redshifts as high as $z \simeq 10.6$ \citep{Kokorev_2023, Fan_2023, Greene_2024, Maiolino_2024, Bogdan_2024, Kovács_2024}.

%While the present work focuses on the local BBH population and its dependence on AGN model assumptions, in companion Paper~II we will explore the implications of these high-redshift SMBH populations for the BBH merger rate and for future third-generation GW detectors.

%This manuscript is structured as follows. \mpv{...}

\section{Methods}
Our framework builds upon the semi-analytical AGN disk population model introduced in \citet{Vaccaro_2023}. Below, we summarize the baseline model and describe the modifications implemented in this work.

\subsection{Hierarchical BBH mergers in AGN disks}
We modelled the formation and hierarchical growth of BBHs in AGN disks using the semi-analytical population-synthesis framework implemented in \fastclusterAGN{}. The code follows the evolution of stellar-mass BHs embedded in AGN disks through a timescale-based approach, allowing us to efficiently explore a broad range of parameters. 

\subsubsection{Initial conditions}
For each simulation, we initialized the mass of the SMBH, $\MSMBH$, its Eddington fraction, $\fEdd = \dot{M}_\bullet/\dot{M}_{\rm Edd}$, and the viscosity parameter of the accretion disk, $\alpha$. Here, $\fEdd$ is defined as the ratio between the SMBH accretion rate, $\dot{M}_\bullet$, and the Eddington accretion rate, $\dot{M}_{\rm Edd}$.

We modeled the physical properties of the accretion disk using the widely used steady-state analytic \citet{SG} disk model (hereafter \citetalias{SG}). For each choice of $\MSMBH$, $\fEdd$, and $\alpha$, we solved the corresponding one-dimensional disk equations self-consistently using the \pagn{}\footnote{\pagn{} is publicly available via GitLab following \href{https://gitlab.com/aatrani/pagn}{this link}.} Python module \citep{Gangardt_2024}. This provided radial profiles for the gas surface density, $\Sigma_{\rm g}$, the disk aspect ratio, $h = H/R$, the temperature, $T$, and the sound speed, $c_{\rm s}$. Here, $H$ denotes the vertical scale height of the disk and $R$ its cylindrical radius. Representative profiles are shown in \autoref{app:pagn}. To assess the sensitivity of our results to the assumed disk structure, we also perform a comparison with the \citet{TQM} disk model, presented in \autoref{app:TQM}.

We vary the SMBH mass over the range $\log \MSMBH / \Msun = 5.0 \text{--} 9.0$ and the accretion rate over the range $\fEdd = 0.001 \text{--} 10$, encompassing typical values expected across cosmic time \citep[e.g.,][]{Greene_2007, Trinca_2022}. The disk viscosity for a \citetalias{SG} disk, and more generally for so-called $\alpha$-disk models, is parametrized by coefficient $\alpha \in \quadre{0,\,1}$.
We adopt $\alpha = 0.01$ and $\alpha = 0.1$ to bracket the plausible range of viscosities in AGN disks. Lower values are expected in weakly ionized regions, where MRI-driven turbulence is inefficient \citep[e.g.][]{Hawley_2011}, while higher values typically arise in highly ionized regions \citep{King_2005, Martin_2019}. We are limited by the assumption of a spatially constant $\alpha$ across the disk radius, consistent with the approximation in the \citetalias{SG} model, despite more recent simulations indicating potential radial variation in viscosity \citep[e.g.,][]{Penna_2012}.

We assume the AGN lifetime $\tau$ to be distributed according to a lognormal distribution with mean $ \log{(\tau/{\rm Myr})} = 0.22$ and standard deviation $\sigma_{\tau} = 0.8$, based on observations of quasars proximity effect \citep{Khrykin_quasar_lifetime}. 

We model the stellar environment around the SMBH through an associated nuclear star cluster (NSC), whose mass scales with $\MSMBH$ according to \citet{Graham_Spitler}. We estimate the characteristic size of the NSC using the empirical relation from \citet{Neumayer_2020}. 
Assuming that a fraction $f_{\rm BH}=0.04$ \citep{Bartos_2017} of the stellar mass is in stellar-origin BHs, this allows us to estimate both the total number of BHs interacting with the AGN disk and their cumulative mass \citep[see][eq.~7, for more details]{Vaccaro_2023}.

We then randomly draw a first-generation ($1g$, i.e., stellar-origin) BHs and place them in the disk. 
Their masses, $m_1$, are sampled from a catalog obtained with the population synthesis code \sevn{} \citep{sevn_2017, sevn_2019, sevn_2020, sevn_2023}, using the fiducial model from \citet{sevn_2023} and considering single stellar evolution only. We assume metallicity $Z=0.02$ (i.e., approximately solar), which matches the typical metallicity at the center of massive galaxies \citep{Gallazzi_2008}. 
%% VERSIONE PIÙ PROLISSA: %%copy&paste from paper I
%\sevn{} \citep{sevn_2017, sevn_2019, sevn_2020, sevn_2023}. \sevn{} relies on up-to-date stellar tracks \citep{Bressan_2012, Costa_2019, Nguyen_2022} and models the formation of compact objects by taking into account electron-capture \citep{Giacobbo_2019}, core-collapse \citep{sevn_2023} and pair-instability supernovae \citep{sevn_2020}. In particular, here we used the rapid core-collapse supernova model from \cite{Fryer_2012}, which enforces the existence of a mass gap between the maximum neutron star mass and the minimum BH mass \citep{Oezel_2010}. We used the fiducial model from \citet{sevn_2023} and considered single stellar evolution only. We assumed metallicity $Z=0.02$ (i.e., approximately solar), which matches the typical metallicity at the center of massive galaxies \citep{Gallazzi_2008}. 

We draw the dimensionless spin magnitude, $\chi_1$, from a Maxwellian distribution with one-dimensional root-mean-square $\sigma_\chi =0.05$, truncated at $\chi=1$. We have chosen $\sigma_\chi = 0.05$ because it is reminiscent of the spins inferred from the fourth GW transient catalog (GWTC-4, \citealt{GWTC4_pop}). %%This assumption does not take into account that the BBH population in GWTC-4 likely comes from multiple formation channels, including the AGN disk scenario. Current data are not sufficiently informative to differentiate between formation channels. 
We also consider a  
scenario in which the spin of $1g$ BHs is drawn from a truncated Gaussian distribution centered at $\chi_1=0$. The results of this additional model are presented in \autoref{app:chi_zero}.

Gas torques in thin, weakly magnetized accretion disks tend to align a BH spin vector, $\vec{\chi}_{1}$, with the angular momentum of the disk, $\vec{J}$ \citep{Dhruv_2025}. We assign nearly aligned spins by drawing the cosine of the angle between $\vec{\chi}_1$ and $\vec{J}$ from a truncated Gaussian centered on $1$, with width $\sigma=0.1$.

The initial radial position of $1g$ BHs is sampled from the steady-state distribution of stellar-mass BHs in a NSC derived by \citet{Rom_2024} (their Eq.~15), which approximates the distribution of BHs around the SMBH prior to capture by the AGN disk. We sample this distribution between the inner and outer disk radii, $R_{\rm min}=3\,R_{\rm S}=6\,G\MSMBH/c^2$ and $R_{\rm max}=\SI{0.1}{\parsec}\,(M_{\rm SMBH}/10^6\Msun)^{1/2}$ \citep{Goodman_2003, Yang_2019_bis}.

\subsubsection{Gas capture, migration, and binary pairing}
The orbital evolution of individual BHs is followed semi-analytically. In particular, gas capture is modeled following \citet{Rowan_2025}, as described in \autoref{app:gas_capture}.

Once embedded in the disk, BHs migrate radially on a timescale $t_{\rm migr}$ due to torques exerted by the surrounding gas. We include both Type I and Type II migration regimes. In the low-mass regime, where the perturber does not significantly modify the disk, we adopt the migration prescriptions from \citet{Grishin_2024}. For more massive BHs, capable of partially opening a gap in the gas disk, we adopt the modified Type II prescription from \citet{Kanagawa_2018}. We compute in detail whether migration can bring BHs within one mutual Hill radius of each other, on which timescale, and at which radial location as in \citet[][see their Appendix B]{Vaccaro_2025}. 

Our treatment of BBH pair-up is based on a gas-assisted pairing criterion \citep{Qian_2024} together with a phenomenological parameter, $f_{\rm progr}$, describing the prograde fraction of newly formed binaries; details are given in \autoref{app:pairing}.

%%%%%%%%%%%%%%%%%%%%%%%%%%%%%%%%%%%%%%%%%%%%%%%%%%%%%%%%%%%%%%%%%%%%%%%%%%%%%%%%%%%
%%%%%%%%%%%%%%%%%%%%%%%%%%%%%%%% FIGURES %%%%%%%%%%%%%%%%%%%%%%%%%%%%%%%%%%%%%%%%%%
%%%%%%%%%%%%%%%%%%%%%%%%%%%%%%%%%%%%%%%%%%%%%%%%%%%%%%%%%%%%%%%%%%%%%%%%%%%%%%%%%%%

\begin{figure*}
    \centering
    \includegraphics[width=\linewidth]{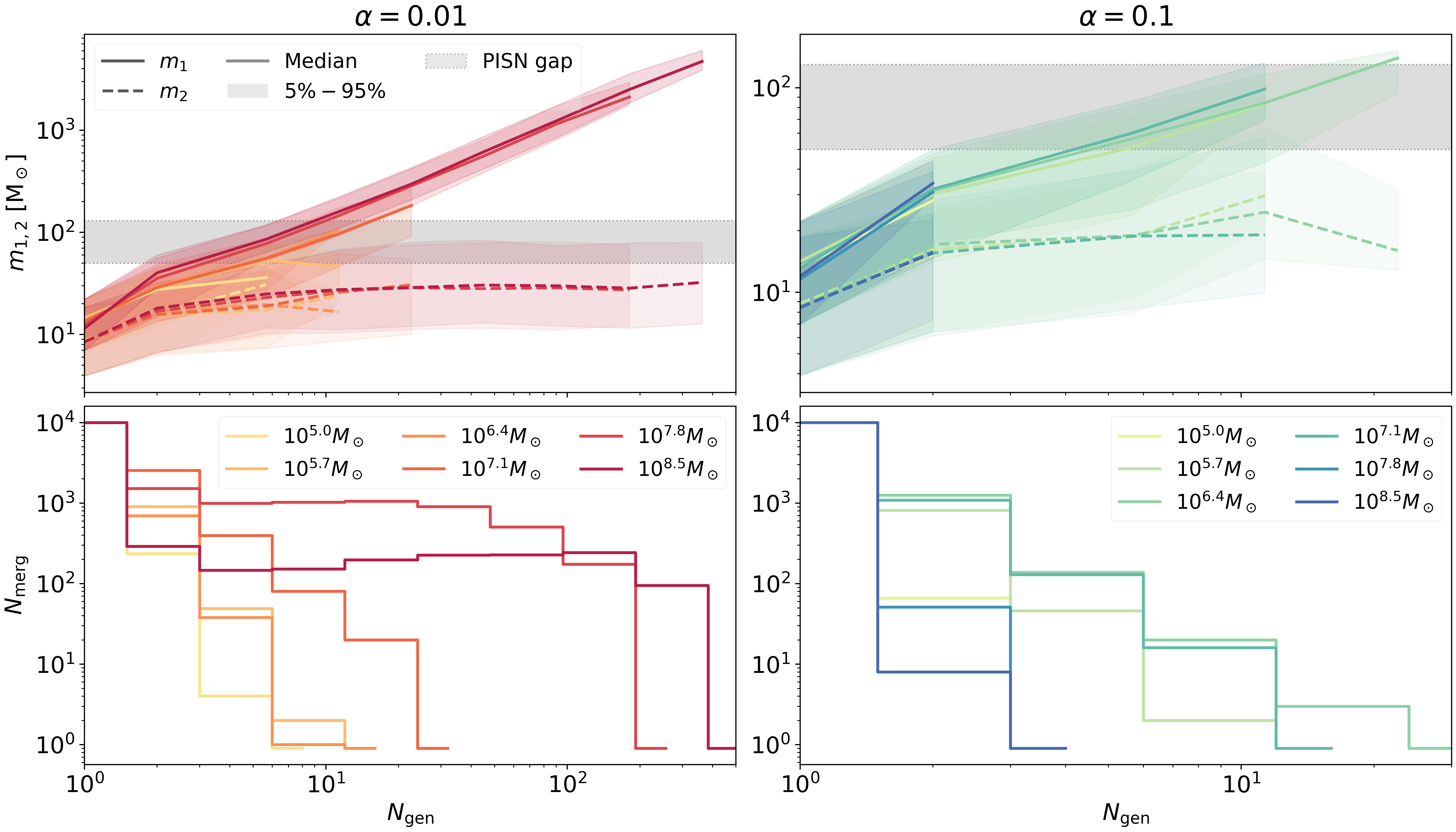}
    \caption{Top panels show the evolution of component masses, $m_{1,2}$, %and dimensionless spins, $\chi_{1,2}$, 
    as a function of merger generation number, $N_{\rm gen}$, for different SMBH masses (color-coded), at fixed $\fEdd=0.1$ for $f_{\rm progr}=1$. Left (right) panels correspond to disks with viscosity parameter $\alpha=0.01$ ($0.1$). Solid (dashed) lines denote the primary (secondary) BH, with shaded regions indicating the $5$–$95$ percentile range. We mark the PISN mass gap, $50 \lesssim m_1/\Msun \lesssim 130$, as predicted for BHs formed through standard single-star evolution \citep{sevn_2017,Woosley_2021} in gray. The lower panels show the number of mergers as a function of generation number.}
    \label{fig:percentiles_vs_ngen}
\end{figure*}

%%%%%%%%%%%%%%%%%%%%%%%%%%%%%%%%%%%%%%%%%%%%%%%%%%%%%%%%%%%%%%%%%%%%%%%%%%%%%%%%%%%
%%%%%%%%%%%%%%%%%%%%%%%%%%%%%%%%%%%%%%%%%%%%%%%%%%%%%%%%%%%%%%%%%%%%%%%%%%%%%%%%%%%
%%%%%%%%%%%%%%%%%%%%%%%%%%%%%%%%%%%%%%%%%%%%%%%%%%%%%%%%%%%%%%%%%%%%%%%%%%%%%%%%%%%

\subsubsection{Binary orbital evolution}

Once a BBH forms, we evolve it in detail under the combined effect of gas hardening \citep{Ishibashi_2020, Ishibashi_2024}, dynamical perturbations from three-body encounters with single BHs \citep{Trani_2024}, and GW emission.

We model the interaction between the BBH and its circumbinary disk following the analytical framework of \citet{Ishibashi_2024}, calibrated on hydrodynamical simulations of circumbinary accretion \citep{Tiede_2020, Heath_2020}. In this picture, gas torques can either shrink or widen the binary depending on the local disk properties, and GW emission becomes dominant at small separations. We therefore evolve the BBH semi-major axis, $a$, and eccentricity, $e$, by combining gas-driven hardening with the standard GW-driven evolution equations of \citet{Peters_1964},
\begin{equation}
    \dot{a}=\dot{a}_\mathrm{g} + \dot{a}_\mathrm{GW}\,, \qquad 
    \dot{e}=\dot{e}_\mathrm{g} + \dot{e}_\mathrm{GW} .
    \label{eq:IG and Peters}
\end{equation}
The complete %explicit 
form of the gas-driven evolution equations, $\dot{a}_\mathrm{g}$ and $\dot{e}_\mathrm{g}$, is given in \autoref{app:binary_evolution}.
We integrate the hardening equations using the Euler method and an adaptive timestep \citep{fastcluster2021} and, if the BBH reaches merger,\footnote{We assumed that the BBH merges when its members cross the ISCO radius of a non-spinning BH with mass equal to the total mass of the binary system, $r_{\rm ISCO}=6\,G\,(m_1+m_2)/c^2$, with a tolerance of $0.1\,r_{\rm ISCO}$.} we record the elapsed time as $t_\mathrm{insp}^{\rm (gas)}$. We also track the eccentricity of the binary when the GW frequency reaches $10\,\mathrm{Hz}$, defining the characteristic GW frequency as the harmonic carrying the largest fraction of the emitted power as in \citet{Wen_2003}.

In addition to gas hardening, BBHs can undergo encounters with other single BHs orbiting in the disk. These encounters  perturb the binary orbit before the binary has time to merge through gas hardening and GW emission alone.

We compare the characteristic encounter timescale, $t_\mathrm{enc}$, to the gas-driven inspiral timescale, $t_\mathrm{insp}^{\rm (gas)}$. If $t_\mathrm{enc} > t_\mathrm{insp}^{\rm (gas)}$, we assume that the BBH evolves uninterrupted until merger. Otherwise, we model one or more encounters using a pre-computed grid of post-Newtonian three-body scattering experiments performed with \tsunami{} \citep{tsunami}. After each surviving encounter, we update the BBH orbital parameters and re-integrate its inspiral using eq.~\ref{eq:IG and Peters}. This procedure is repeated until the BBH merges, is disrupted, or the cumulative evolution time exceeds the AGN lifetime. Further details are given in \autoref{app:binary_evolution} and \autoref{app:3bb}.
The total inspiral time is therefore
\begin{equation}
t_{\rm insp} = t_{\rm insp}^{\rm (gas)} + N_{\rm enc} \, t_{\rm enc},
\label{eq:inspiral_timescale}
\end{equation}
where $t_{\rm insp}^{\rm (gas)}$ is the gas-driven inspiral timescale after the latest encounter, and $N_{\rm enc}$ is the number of encounters recorded for the BBH before it reaches its merger condition.
This treatment effectively captures the interplay between smooth gas-driven inspiral and stochastic perturbations from repeated encounters in the AGN disk environment.

\subsubsection{Nth-generation (\texorpdfstring{$N$g}{Ng}) BHs}
\label{sec:Nth_gen_model}

If a BBH merges, we compute the remnant mass and the aligned component of the remnant spin using the aligned-spin prescription of \citet{Jimenez}. Since this prescription does not include precessing configurations, we supplement it with the correction proposed by \citet{JohnsonMcDaniel_2016}. The recoil kick velocity is computed following \citet[][their eq.~14.202]{maggiore_GW}. The remnant BH can then remain bound to the AGN disk and participate in subsequent mergers, allowing for the formation of hierarchical merger chains.

We follow the hierarchical evolution of each BH using the same semi-analytical implementation adopted for $1g$ BHs, until one of the following stopping conditions is met. 
First, the cumulative merger timescale may exceed the AGN lifetime,
\begin{equation}
t_\mathrm{merg}^{(N)} = t_\mathrm{in}^{(N)} + \left(t_\mathrm{damp} + t_\mathrm{migr} + t_\mathrm{pair} + t_\mathrm{insp}\right)^{(N)} > \tau,
\end{equation}
where $t_\mathrm{in}^{(N)}$ accounts for the time spent in previous generations.

Second, the recoil kick imparted during the merger may eject the remnant from the AGN disk if its post-merger velocity exceeds the local escape velocity,
$\left|\vec{v}_\mathrm{K} + \vec{v}_\mathrm{kick}\right| > v_\mathrm{esc}(R)$.
Finally, hierarchical growth may stop simply because the BH has exhausted the available supply of companions.%, namely
%\begin{equation}
%M_\mathrm{acc} = m_1 + \sum_N m_2^{(N)} > M_\mathrm{BH}^\mathrm{max}.
%\end{equation}
Additional details on the implementation of hierarchical mergers are provided by \citet{Vaccaro_2023}.

\subsection{Intrinsic BBH mergers population}\label{sec:cosmo_methods}

To construct an intrinsic BBH merger population at a given redshift, we combine the outputs of all our \fastclusterAGN{} runs, each corresponding to a different pair of SMBH mass and accretion rate, $(\MSMBH, \fEdd)$, assigning to each run a weight that reflects the expected distribution of AGNs at the chosen cosmic epoch.

For a given redshift $z$, we draw SMBH masses from a probability distribution $p(\MSMBH\mid z)$ and, at fixed SMBH mass, Eddington ratios from a conditional distribution $p(\fEdd\mid \MSMBH, z)$. The weight associated with a run characterized by $(M_{\bullet i}, \lambda_{\bullet j})$ is then proportional to
\begin{equation}
w_{ij}(z) \propto p(M_{\bullet i}\mid z)\, p(\lambda_{\bullet j}\mid M_{\bullet,i}, z)\, \Delta M_{\bullet i}\, \Delta\lambda_{\bullet j},
\end{equation}
where $\Delta M_{\bullet i}$ and $\Delta \lambda_{\bullet j}$ are the effective widths of the corresponding grid bins. We normalize the weights such that $\sum_{i,j} w_{ij}(z) = 1$. 

Quantities such as the BBH mass and spin distributions are then computed by combining the contribution of each run weighted by $w_{ij}(z)$.
More generally, the probability density of an observable $x$ at redshift $z$ can be written as
\begin{equation}
p(x\mid z) = \sum_{i,j} w_{ij}(z)\, p(x\mid M_{\bullet i}, \lambda_{\bullet j}),
\end{equation}
where $p(x\mid M_{\bullet i}, \lambda_{\bullet j})$ is estimated directly from the BBH mergers produced in the corresponding $(\MSMBH, \fEdd)$ \fastclusterAGN{} realization. %\new{With this weighting procedure we construct an episode-integrated merger population, i.e. the distribution of BBH properties among mergers produced during AGN episodes drawn from an adopted AGN population. We do not intend this as an absolute merger-rate density, which is deferred to a companion paper.}

Here, we adopt the AGN population properties from \citet{Greene_2007}, which are derived from SDSS broad-line AGNs and are representative of the local Universe ($z \lesssim 0.3$). Therefore, we sample $\log(\MSMBH/\Msun)$ from a normal distribution with mean $\mu=6.576$ and standard deviation $\sigma=0.591$, and draw the $\log\fEdd$ from a normal distribution with mean $\mu=-1$ and standard deviation $\sigma=0.25$.  
The resulting weighted mixture provides an intrinsic BBH merger population representative of the AGN channel at low redshift.

\subsection{Description of runs}
We ran $10^4$ independent Monte Carlo realizations of our semi-analytical model for each combination of $\MSMBH$, $\fEdd$, $\alpha$ and $f_{\rm prog}$, generating a large ensemble of BBH mergers that captures the stochastic nature of the underlying processes. We sample over a grid $\log (\MSMBH / \Msun) \in [5, 9]$ with a uniform spacing of $0.1$ dex, and $\log \fEdd \in [-3, 1]$ in steps of $1$ dex. We consider two values of viscosity parameters, $\alpha = 0.01, 0.1$ and three values of $f_{\rm progr}=0, 0.5, 1$. %The results presented in \autoref{sec:param_space} are based on these suit of realizations and are used to map the efficiency of BBH formation and hierarchical growth across the AGN parameter space. 
We weight each realization according to $p(\MSMBH|z)$ and $p(\fEdd|\MSMBH, z)$ in the local Universe \citep{Greene_2007}, thereby building a mixture of BBH populations across multiple AGN systems.%, presented in \autoref{sec:local_pop}.}

%To construct a BBH merger population representative of the local Universe, we combine the outputs of all simulations by assigning a weight to each $(\MSMBH, \fEdd)$ configuration based on the expected distribution of AGN properties at low redshift (\citealt{Greene_2007}; see Section~\ref{sec:cosmo_methods}). In practice, this corresponds to weighting each realization according to $p(\MSMBH|z)$ and $p(\fEdd|\MSMBH, z)$, thereby building a mixture of BBH populations across the parameter grid. %This procedure effectively maps the grid of AGN environments onto a population-level prediction for BBH mergers in the local Universe.

%%%%%%%%%%%%%%%%%%%%%%%%%%%%%%%%%%%%%%%%%%%%%%%%%%%%%%%%%%%%%%%%%%%%%%%%%%%%%%%%%%%
%%%%%%%%%%%%%%%%%%%%%%%%%%%%%%%% FIGURES %%%%%%%%%%%%%%%%%%%%%%%%%%%%%%%%%%%%%%%%%%
%%%%%%%%%%%%%%%%%%%%%%%%%%%%%%%%%%%%%%%%%%%%%%%%%%%%%%%%%%%%%%%%%%%%%%%%%%%%%%%%%%%

\begin{figure*}
    \centering
    \includegraphics[width=\linewidth]{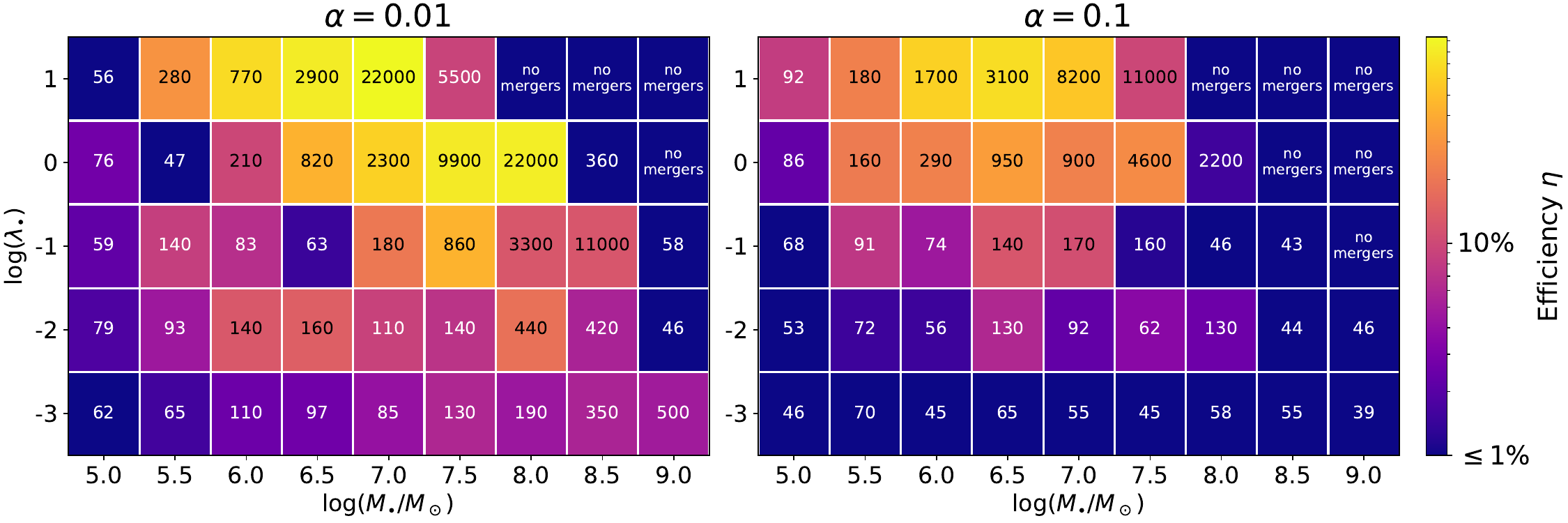}
    \caption{Each panel shows the dependence of BBH merger outcomes on SMBH mass, $\MSMBH$ (x-axis), and Eddington ratio, $\fEdd$ (y-axis), for $f_{\rm progr}=0.5$ two values of the disk viscosity parameter: $\alpha=0.01$ (left) and $\alpha=0.1$ (right). The %top panels 
    labels report the maximum merger-product mass (in $\Msun$) achieved in each configuration, while the %bottom panels 
    colorbar show the corresponding merger efficiency, $\eta=N_{\rm merg}/N_{\rm tot}$. Values are indicated in each cell, bins with no mergers are labeled accordingly.}
    \label{fig:max_mbinary_eta_grid}
\end{figure*}

\begin{figure*}
    \centering
    \includegraphics[width=\linewidth]{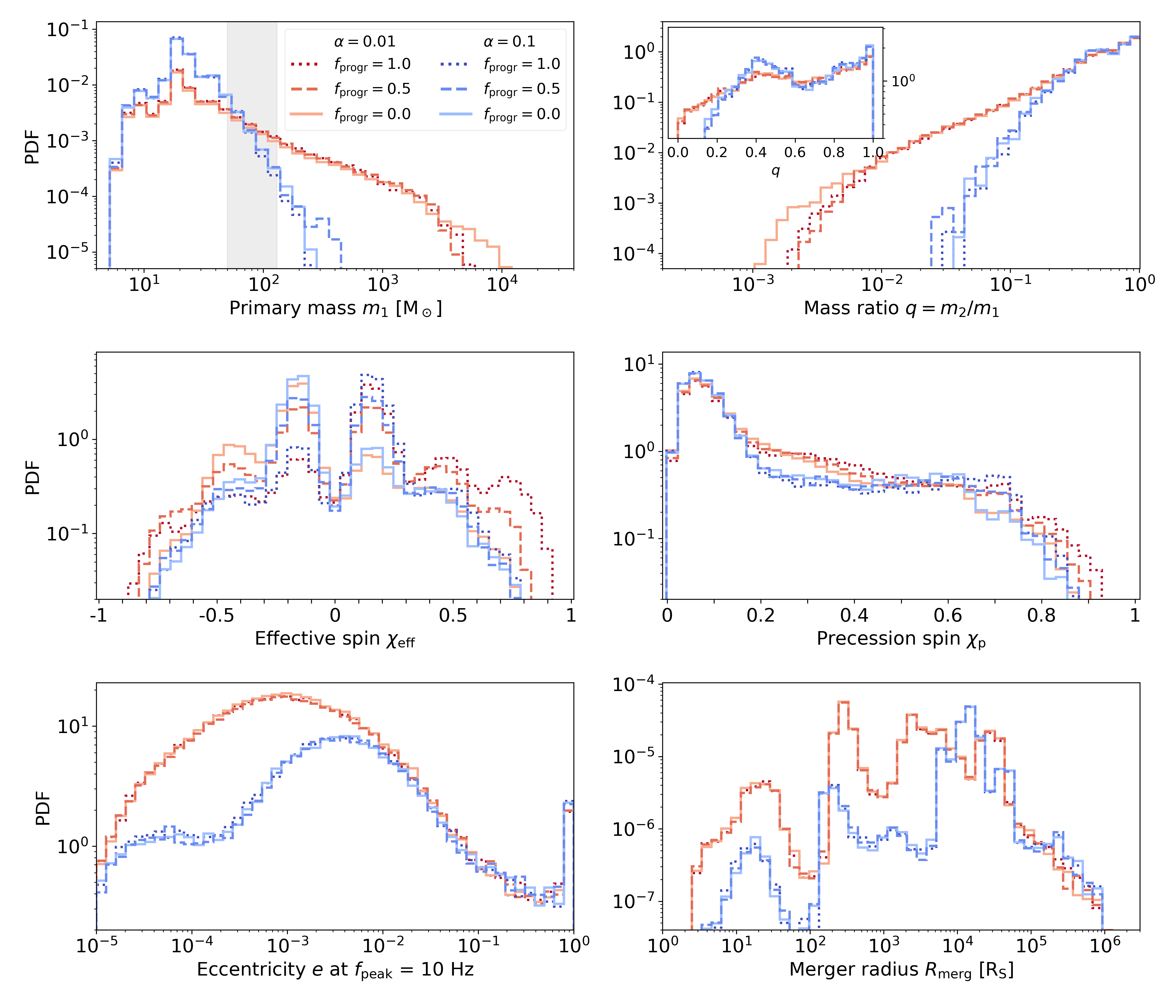}
    \caption{Probability density functions of the properties of BBH mergers in the local AGN population \citep{Greene_2007}, for different values of the disk viscosity parameter, $\alpha$, and fraction of prograde BBHs at formation, $f_{\rm progr}$. From top left to bottom right: primary mass $m_1$, mass ratio $q$, effective spin $\chi_{\rm eff}$, precession spin $\chi_p$, eccentricity at $f_{\rm peak} = 10$ Hz, and merger radius $R_{\rm merg}$ in units of Schwarzschild radii. We mark the PISN mass gap, $50 \lesssim m_1/\Msun \lesssim 130$, as predicted for BHs formed through standard single-star evolution \citep{sevn_2017,Woosley_2021} in gray. Line colors encode the viscosity parameter, $\alpha$, while line styles correspond to different orbital configurations ($f_{\rm progr}$), ranging from fully prograde to fully retrograde.}
    \label{fig:weighted_mixture_local_histograms}
\end{figure*}

\begin{figure*}
    %\sidecaption
    %\includegraphics[width=12cm]{figures/weighted_mixture_local_corner_contours_binaries_bigger.png}
    \centering
    \includegraphics[width=0.85\linewidth]{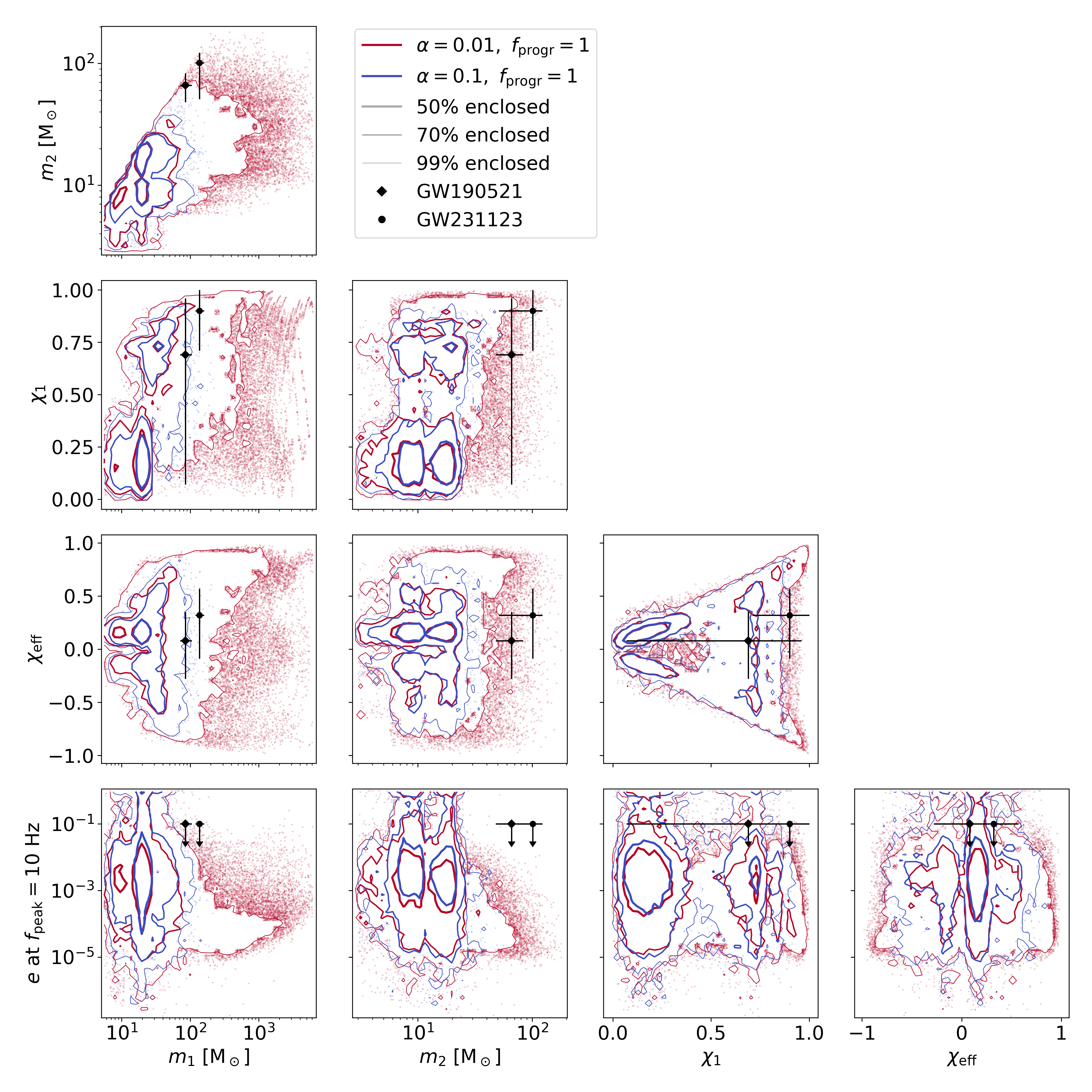}
    \caption{Distributions of BBH properties for the local AGN population, comparing two disk viscosities: $\alpha=0.01$ (red) and $\alpha=0.1$ (blue), both for fully prograde configurations ($f_{\rm progr}=1$). Panels show the joint distributions of primary mass $m_1$, secondary mass $m_2$, primary spin $\chi_1$, effective spin $\chi_{\rm eff}$, and eccentricity $e$ at 10 Hz. Contours enclose $50\%$, $70\%$, and $99\%$ of the probability density,while points denote systems lying outside the corresponding $99\%$ contour. Black markers indicate the GW events GW190521 and GW231123, with error bars showing their observational uncertainties \citep{GW190521, GW231123}.}
    \label{fig:weighted_mixture_local_corner_contours_binaries}
\end{figure*}

%%%%%%%%%%%%%%%%%%%%%%%%%%%%%%%%%%%%%%%%%%%%%%%%%%%%%%%%%%%%%%%%%%%%%%%%%%%%%%%%%%%
%%%%%%%%%%%%%%%%%%%%%%%%%%%%%%%%%%%%%%%%%%%%%%%%%%%%%%%%%%%%%%%%%%%%%%%%%%%%%%%%%%%
%%%%%%%%%%%%%%%%%%%%%%%%%%%%%%%%%%%%%%%%%%%%%%%%%%%%%%%%%%%%%%%%%%%%%%%%%%%%%%%%%%%

\section{Results}
\label{sec:results}

%\subsection{Parameter space exploration}\label{sec:param_space}

\subsection{The pair instability mass gap shows up in the secondary mass function}
We explore how the properties of BBH mergers depend on the main parameters of the AGN environment, and provide a global view of hierarchical growth across the AGN parameter space. 
\autoref{fig:percentiles_vs_ngen} shows the evolution of BBH properties as a function of generation number, $N_{\rm gen}$, for selected values of $\MSMBH$, fixed $\fEdd=0.1$, and for both values of $\alpha=0.01, 0.1$. The number of mergers decreases steeply with generation number. High-generation mergers are therefore rare even in the most favorable regions of parameter space, indicating that very massive BHs produced through repeated mergers populate the tail of the distribution.

Across all configurations, the primary BH mass increases monotonically with generation number, while the secondary mass evolves more weakly and saturates at lower values. This behavior is a direct consequence of  how we assign the generation of the companion: %For a primary of generation $N$, the secondary generation $M$ is drawn from a distribution, $p(M)\propto 2^{-(M-1)}$ with $M\leq N$ (see \autoref{app:pairing}), which reflects the decreasing abundance of higher-generation BHs in the disk. 
we select the properties of the secondary BH to reflect the decreasing abundance of higher-generation BHs in the disk, so that high-generation primaries thus typically merge with lower-generation secondaries rather than forming equal-generation $N$g--$N$g binaries. %This is a distinctive feature of our AGN-channel model: unlike environments where mass segregation can preferentially pair the most massive objects with each other, the AGN channel favors mergers between a hierarchically grown primary and a lower-generation companion.
As a result, hierarchical growth efficiently populates, and eventually smooths out, the pair-instability supernova (PISN) mass gap in the primary mass distribution, whereas secondary BHs rarely exceed $\sim 50\Msun$, remaining below the gap, because they are usually drawn from lower hierarchical-merger generations. The population therefore evolves toward increasingly unequal-mass systems at high generation number. This behavior is qualitatively consistent with recent population-level evidence that the pair-instability gap may more clearly show up in the $m_2$ distribution than in the $m_1$ distribution \citep{Tong_2026}.

%\subsection{\new{Impact of the viscosity parameter $\alpha$}}
For $\alpha=0.01$, mass growth is more sustained than for higher viscosity, reaching high merger generations ($N_{\rm gen}>100$) and producing more massive binaries, particularly for $\MSMBH\sim10^8 \Msun$. In contrast, for $\alpha=0.1$, hierarchical growth is less efficient and typically saturates at lower generations ($N_{\rm gen}\simeq20$).
This comparison shows that the growth history is highly sensitive to the underlying disk structure. As shown in \autoref{fig:pAGN_profiles}, at fixed $\MSMBH$ and $\fEdd$, lower-viscosity disks ($\alpha=0.01$) have higher gas surface densities than disks with larger $\alpha$ ($\alpha=0.1$). This in turn affects the disk-driven processes that regulate binary formation and evolution. In the low-$\alpha$ models, the larger $\Sigma_{\rm g}$ shortens several relevant dynamical timescales, including those associated with capture (see \autoref{app:gas_capture}), migration \citep{Vaccaro_2025}, pairing (eq.~\ref{eq:t_pair}), and binary-single encounters (eq.~\ref{eq:t_enc}), consequently making hierarchical mergers more efficient.

\subsection{Merger efficiency in AGN disks}
\autoref{fig:max_mbinary_eta_grid} summarizes how the merger population varies across the parameter space. We show both the maximum BBH merger-product mass, $\max(m_{\rm prod})$, defined as the largest BH mass produced by a BBH merger within a given realization, and the merger efficiency, $\eta = N_{\rm merg}/N_{\rm tot}$. These quantities probe complementary aspects of the process: $\max(m_{\rm prod})$ traces the extent of hierarchical growth, while $\eta$ captures the overall level of dynamical activity.

Only a limited region of the parameter space is sufficiently dynamically active to produce both high efficiencies and very massive remnants. Outside of this region, either mergers are rare ($\eta \ll 1$) or hierarchical growth saturates at relatively modest masses. In particular, some configurations at high $\MSMBH$ show no mergers at all, indicating that a large SMBH mass alone does not guarantee efficient BBH assembly.

Across both values of $\alpha$, we find a general increase of $\eta$ with $\fEdd$. A similar, though less uniform, trend is seen in $\max(m_{\rm rem})$. However, this behavior is not monotonic: both quantities tend to peak at intermediate-to-large SMBH masses, with the location of the peak depending on $\fEdd$.
%The comparison between $\alpha=0.01$ and $\alpha=0.1$ highlights the sensitivity of these trends to the underlying disk structure. Lower-viscosity disks generally allow for more sustained hierarchical growth and higher efficiencies. This reflects a delicate interplay between migration, encounter rates, and gas-driven hardening, all of which depend on $\alpha$, $\MSMBH$, $\Sigma_{\rm g}$, $h$, and $c_{\rm s}$. This behavior is consistent with previous findings \citep{Vaccaro_2025}, and highlights that the efficiency of the AGN channel is primarily controlled by the disk-driven dynamical timescales.
Overall, the AGN channel spans a wide diversity of outcomes across the explored parameter space, from nearly inactive systems to environments capable of sustained hierarchical growth, without exhibiting a single characteristic behavior.

\subsection{Local population of BBH mergers from AGN disks}\label{sec:local_pop}

%%%%%%%%%%%%%%%%%%%%%%%%%%%%%%%%%%%%% TABLE %%%%%%%%%%%%%%%%%%%%%%%%%%%%%%%%%%%%% 
\begin{table}
\centering
\caption{
Fraction of simulated local AGN-assisted mergers that are at least as extreme as the median inferred parameters of GW190521 \citep{GW190521} and GW231123 \citep{GW231123} in selected two-dimensional planes of the parameter space. 
}
\label{tab:event_tail_fractions}
\begin{tabular}{cccccc}
\hline 
\hline
$\alpha$ & $f_{\rm prog}$ & Parameters & GW190521 & GW231123 \\
\hline
$0.01$ & 1 
& $(m_1,m_2)$ 
& $2.0\times10^{-3}$  
& $2.9\times10^{-4}$ \\

$0.01$ & 0.5 
& $(m_1,m_2)$ 
& $2.1\times10^{-3}$  
& $3.0\times10^{-4}$  \\

$0.01$ & 0 
& $(m_1,m_2)$ 
& $2.0\times10^{-3}$ 
& $2.9\times10^{-4}$  \\

$0.01$ & 1 
& $(m_1,\chi_{\rm eff})$ 
& $2.1\times10^{-2}$  
& $1.1\times10^{-2}$ \\

$0.01$ & 0.5 
& $(m_1,\chi_{\rm eff})$ 
& $1.6\times10^{-2}$  
& $5.1\times10^{-3}$  \\

$0.01$ & 0 
& $(m_1,\chi_{\rm eff})$ 
& $9.3\times10^{-3}$ 
& $2.3\times10^{-3}$  \\

$0.1$ & 1 
& $(m_1,m_2)$ 
& $1.0\times10^{-4}$  
& $0$ \\

$0.1$ & 0.5 
& $(m_1,m_2)$ 
& $7.8\times10^{-5}$  
& $0$ \\

$0.1$ & 0
& $(m_1,m_2)$ 
& $5.7\times10^{-5}$  
& $0$ \\

$0.1$ & 1
& $(m_1,\chi_{\rm eff})$ 
& $2.7\times10^{-3}$  
& $4.0\times10^{-4}$  \\

$0.1$ & 0.5
& $(m_1,\chi_{\rm eff})$ 
& $3.5\times10^{-3}$ 
& $4.7\times10^{-4}$ \\

$0.1$ & 0 
& $(m_1,\chi_{\rm eff})$ 
& $4.0\times10^{-3}$ 
& $3.5\times10^{-4}$ \\

\hline
\end{tabular}
\end{table}
%%%%%%%%%%%%%%%%%%%%%%%%%%%%%%%%%%%%%%%%%%%%%%%%%%%%%%%%%%%%%%%%%%%%%%%%%% 

We characterize the properties of BBH mergers in the local Universe ($z \approx 0$) arising from AGN disks by combining the populations obtained across the explored $(\MSMBH, \fEdd)$ parameter space according to \citet{Greene_2007}. In this section, we focus on the distributions of key observables and on their dependence on global AGN properties.

\autoref{fig:weighted_mixture_local_histograms} shows the distributions of the main observable properties of BBH mergers in the local AGN population, for different values of the disk viscosity parameter, $\alpha$, and fraction of initially prograde orbits, $f_{\rm progr}$. The primary mass distribution spans a broad range, from stellar-mass BHs to $m_1\gtrsim 10^3 M_\odot$, with a clear high-mass tail produced by hierarchical mergers. This tail is more prominent for lower viscosity ($\alpha=0.01$), while higher viscosity ($\alpha=0.1$) suppresses the formation of the most massive systems. The mass ratio distribution extends toward unequal-mass binaries ($q \ll 1$), especially at lower $\alpha$, and presents peaks at $q\simeq1$ and $q\simeq0.4$. These features are associated to $1$g mergers (see \autoref{app:q_Ngen}) and are
most clearly visible for $\alpha=0.1$, for which hierarchical growth is less efficient, while in the
$\alpha=0.01$ models they are washed out by higher-generation
mergers, which broaden the distribution toward lower mass ratios.

The effective spin, $\chi_{\rm eff}$, spans the full allowed range and encodes the degree of orbital alignment set by the AGN disk. Its detailed shape depends strongly on the fraction of binaries that are prograde at birth, $f_{\rm progr}$. For a $f_{\rm progr}= 1$, $\chi_{\rm eff}$ is skewed toward positive values, reflecting the preferential alignment of spins with the orbital angular momentum. Conversely, for $f_{\rm progr}=0$, the distribution shifts toward negative $\chi_{\rm eff}$. However, the sign of $\chi_{\rm eff}$ is not uniquely determined by the initial prograde or retrograde configuration: three-body encounters can tilt the orbital plane and modify the relative orientation between spins and orbital angular momentum. As a result, even binaries that are initially prograde (retrograde) can contribute to the negative (positive) tail of the $\chi_{\rm eff}$ distribution.
For intermediate values, and in particular $f_{\rm progr}=0.5$, the distribution becomes clearly bimodal, with symmetric peaks at positive and negative $\chi_{\rm eff}$. However, this bimodality wildly depends on first-generation spin magnitudes, as it fully disappears if we assume nearly non-spinning stellar-origin BHs, as shown in \autoref{app:chi_zero}. The sharpness of the effective spin features depends on the contribution of higher-generation mergers, which tend to reinforce coherent spin orientations. Indeed, in the lower-viscosity case, $\alpha=0.01$, the 
peaks at $\left| \chi_{\rm eff} \right|\simeq0.4$ and $\chi_{\rm eff} \simeq0.8$ are more pronounced than in the $\alpha=0.1$ case.
The precession spin, $\chi_p$, shows a peak at $\chi_p \simeq 0.1$ with a tail extending to $\chi_p \sim 1$, indicating that while most binaries remain moderately aligned, a non-negligible fraction retains significant in-plane spin components. 
Recent population studies have reported evidence for a distinct subpopulation of massive, rapidly spinning BBHs whose spins are preferentially aligned with the orbital angular momentum \citep{Li_2024, Li_2025, Rinaldi_2026, Tong_2026, Antonini_2025, Flanagan_2026, Hussain_2026, Padhyegurjar_2026}. These systems qualitatively resemble the high-mass, positive-$\chi_{\rm eff}$ population produced by hierarchical mergers in our AGN disk models, although it is not possible to exclude other interpretations, given observational and theoretical uncertainties.

The eccentricity at $10$ Hz is generally low, peaking at $e \sim 10^{-3}$ ($10^{-2}$) for $\alpha=0.01$ ($0.1$), but also exhibits a secondary peak at $e \gtrsim 0.8$. This bimodality indicates that while most binaries efficiently circularize through GW emission before entering the detector band, a subset enters the observational window earlier in its evolution, before significant circularization has taken place, and therefore retains the imprint of dynamical interactions and gas-driven hardening.
The merger radius spans several orders of magnitude, from a few to $\sim 10^6 R_{\rm S}$, with a multi-peaked structure reflecting the different pair-up regions corresponding either to migration traps or traffic jam accumulations\footnote{Traffic jams correspond to overdensities of pair-ups at radii that are not formal migration traps, but where sharp variations in the slope of the torque profile, $\Gamma$, slow down or concentrate migrating BHs.} \citep{Vaccaro_2025}. Most notably, the low-viscosity case ($\alpha=0.01$) shows a pronounced excess of mergers in the inner disk, at $R \simeq 30 R_{\rm S}$.

%Moreover, 
As displayed in \autoref{fig:weighted_mixture_local_corner_contours_binaries}, the AGN channel exhibits distinctive correlations among BBH properties, arising from the interplay between hierarchical mergers and disk-driven dynamics. As the primary mass, $m_1$, grows through successive mergers, binaries progressively shift toward lower mass ratios, $q = m_2/m_1$. This reflects the fact that hierarchical growth in AGNs preferentially increases the mass of the most massive object, while the secondary mass saturates more rapidly, leading to increasingly unequal-mass systems. %\new{This differs from other dynamical channels, where the most massive objects tend to pair up with each other due to mass segregation, producing a stronger preference for nearly equal-mass binaries with $q\simeq 1$ \citep[e.g.][]{Torniamenti_2024}.}

The magnitudes of the primary spin, $\chi_1$, and of the effective spin, $|\chi_{\rm eff}|$, also increase with $m_1$. This trend is intrinsic to the AGN channel and originates from the preferential alignment of angular momenta within the disk: repeated mergers tend to reinforce coherent spin orientations, progressively driving $\chi_{\rm eff}$ toward more extreme values. 
%While this correlation is not strictly deterministic at low generation number, where dynamical interactions and three-body encounters can perturb the orbital plane and broaden the $\chi_{\rm eff}$ distribution, their impact becomes subdominant at $N_{\rm gen} \gtrsim 10$. As a result, the alignment-driven growth of $|\chi_{\rm eff}|$ becomes increasingly efficient at higher masses, and the correlation between $m_1$ and $|\chi_{\rm eff}|$ correspondingly tightens along the hierarchical merger sequence.
The eccentricity at $10$ Hz shows an opposite dependence on mass. Only binaries with $m_1 \sim 20 M_\odot$ retain appreciable eccentricity in the LVK band, whereas more massive systems are nearly circular, with characteristic eccentricities $e<10^{-2}$ for $f_{\rm peak}\gtrsim10$ Hz.

\subsection{GW190521 and GW231123}

%Finally, w
We compare our synthetic population with two BBH merger events in the GWTC-4 catalog \citep{GWTC4_pop} whose remnants lie in the IMBH mass range: GW190521 \citep{GW190521} and GW231123 \citep{GW231123}. We find that AGN disks can in principle produce systems with properties comparable to both events, although these systems occupy the tails of the predicted distributions.

We compute the fraction of simulated mergers in the local population that are at least as extreme as the median inferred event parameters in selected two-dimensional planes. The results are summarized in \autoref{tab:event_tail_fractions}. In the $(m_1,m_2)$ plane, GW190521-like systems occur in the lower-viscosity models ($\alpha=0.01$) at the level of $2\times10^{-3}$, corresponding to roughly one in $500$ mergers. GW231123-like systems are rarer, with fractions of $\sim 3\times10^{-4}$, or roughly one in $3000$ mergers. These fractions are substantially smaller in the higher-viscosity models ($\alpha=0.1$): GW190521-like systems occur only every $10^4$ mergers, while no simulated merger satisfies the corresponding two-dimensional exceedance criterion for GW231123.
The comparison in the $(m_1,\chi_{\rm eff})$ plane leads to a less severe tension. In this projection, the tail fractions are generally larger and depend more strongly on the assumed prograde fraction. Specifically, for $f_{\rm progr}=0.5$, GW190521-like events happen every $60$ ($300$) mergers in the $\alpha=0.01$ ($0.1$) case, whereas GW231123-like events happen every $200$ ($2000$) mergers for $\alpha=0.01$ ($0.1$).

The comparison therefore suggests that the AGN channel, as modeled here, tends to underpredict the secondary mass compared to the values inferred for GW190521 and GW231123. While the corresponding $m_1$ and $\chi_{\rm eff}$ can be obtained in the simulated AGN-disk population at non-negligible rates, especially for $\alpha=0.01$, systems that also match $m_2$ are substantially rarer.
This is particularly the case for GW231123, whose component masses may be more readily reproduced in other scenarios, such as isolated evolution of Population III stars \citep{Tanikawa_2026} or dynamics in low-metallicity stellar clusters \citep{Paiella_2025, Liu_2025}. However, the level of agreement is channel-dependent: the isolated Population III scenario may provide a reasonable match to both masses and spins, whereas the cluster interpretation appears more successful for the component masses than for the spin properties.
Overall, this comparison suggests that the origin of massive BBH merger events remains difficult to constrain: different channels may reproduce different subsets of their inferred properties, and a systematic multi-channel interpretation is left for future work.

%In particular, GW231123 occupies a region where the model underpredicts the secondary mass, highlighting a deficit of high-$m_2$ systems in our population. 
%This suggests that such events are possible but intrinsically rare within the AGN channel as modeled here. 

\section{Discussion}

\subsection{The effect of the AGN lifetime}

We assess the impact of the AGN lifetime by imposing an upper cutoff on the merger time in post-processing. Specifically, we select only mergers with $t_{\rm merg}<\tau_{\rm max}$, and vary the duration of a single AGN episode over the range $\tau_{\rm max}=0.1$--$100$ Myr.

As expected, longer AGN lifetimes allow merger chains to proceed to higher generations, producing a progressively more extended high-mass tail. In particular, the formation of IMBHs with $m_1 \gtrsim 100\,\Msun$ ($1000\,\Msun$) requires $\tau_{\rm max} \gtrsim 1$ Myr ($10$ Myr), as shorter episodes truncate the chains before repeated mergers can efficiently build up massive remnants.

The lifetime distribution adopted here is taken from \citet{Khrykin_quasar_lifetime}, inferred from quasar proximity zones at $2.7<z<3.9$. This is the relevant timescale for our calculation, since proximity zones probe the duration of a single accretion episode rather than the cumulative AGN duty cycle. However, this prescription does not encode possible correlations between $\tau$, $\MSMBH$, and $\fEdd$. Existing constraints on such correlations remain limited: \citet{Worseck_2021} do not find clear evidence for a dependence on either $\MSMBH$ or $\fEdd$, although their sample is restricted to luminous quasars with $\log \MSMBH \gtrsim 8.5$ and $\fEdd \in [0.2, 3]$.

Overall, the AGN lifetime emerges as a key parameter controlling the efficiency of hierarchical growth. In our framework, longer accretion episodes directly translate into a higher probability of forming massive remnants, including IMBHs. If correlations between $\tau$ and $\MSMBH$, $\fEdd$, $\alpha$, or redshift are present, they could significantly reshape the resulting BBH population by enhancing or suppressing hierarchical growth in specific regions of parameter space.
Finally, if %AGN activity is episodic, with 
a galaxy undergoes multiple distinct AGN accretion phases over %the 
its lifetime, %of a galaxy, 
the cumulative dynamical output may be higher than what is predicted for a single AGN episode. 

%%%%%%%%%%%%%%%%%%%%%%%%%%%%%%%%%%%%%%%%%%%%%%%%%%%%%%%%%%%%%%%%%%%%%%%%%%%%%%%%%%%
%%%%%%%%%%%%%%%%%%%%%%%%%%%%%%%% FIGURES %%%%%%%%%%%%%%%%%%%%%%%%%%%%%%%%%%%%%%%%%%
%%%%%%%%%%%%%%%%%%%%%%%%%%%%%%%%%%%%%%%%%%%%%%%%%%%%%%%%%%%%%%%%%%%%%%%%%%%%%%%%%%%

\begin{figure}
    \centering
    \includegraphics[width=\linewidth]{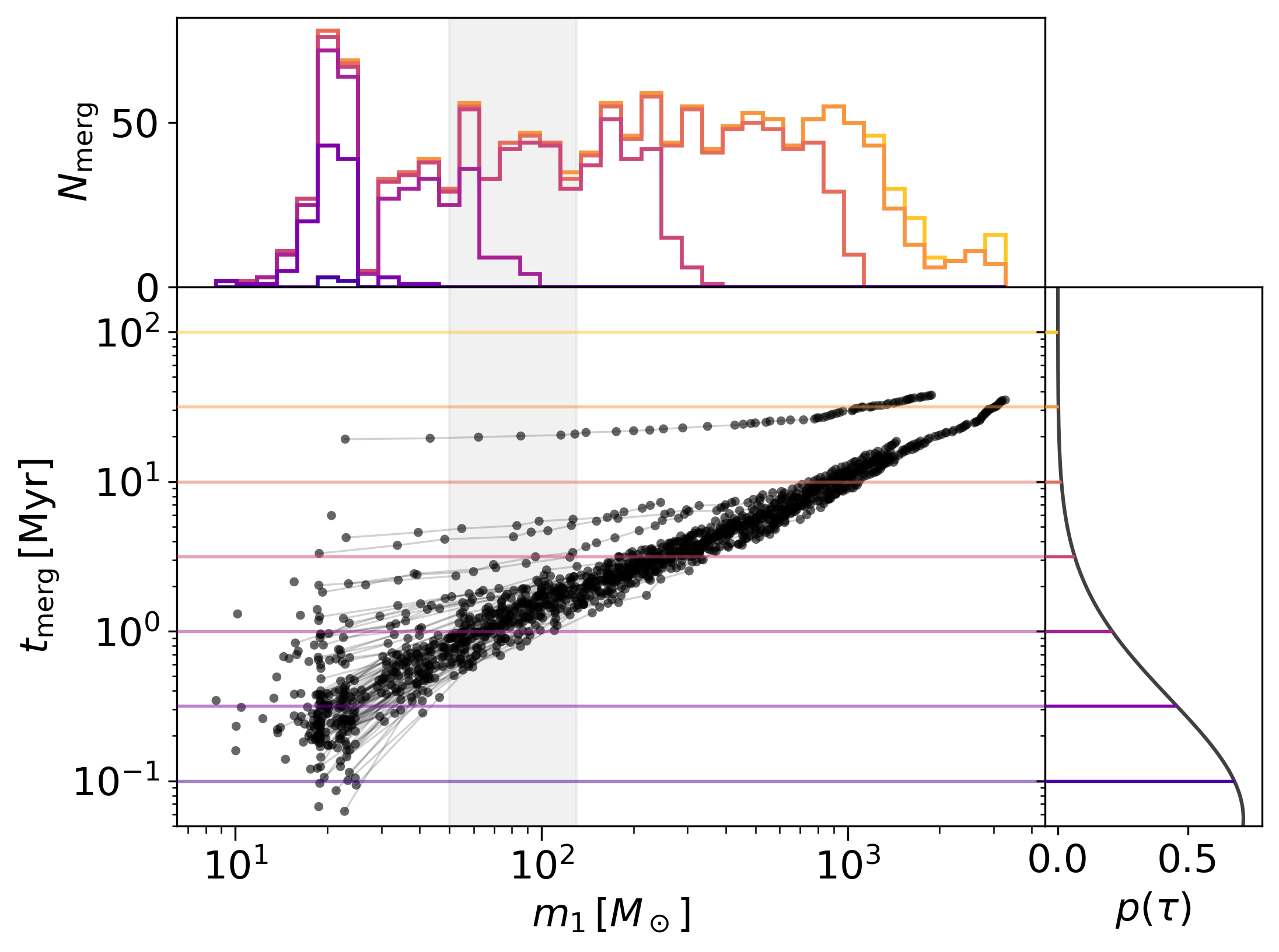}
    \caption{Main panel: distribution of merger timescales, $t_{\rm merg}$, as a function of primary black hole mass, $m_1$, for an AGN system with $\MSMBH=10^8\Msun$, $\fEdd=0.1$ and $\alpha=0.01$. Each point represents an individual merger, while gray line segments connect successive generations within the same hierarchical chain. Right panel: probability density $p(\tau)$ of the AGN lifetime. Colored horizontal lines mark the values of $\tau_{\rm max}$ adopted in the analysis, which truncate the merger chains by selecting events with $t_{\rm merg} < \tau_{\rm max}$. Top panel: distribution of $m_1$, color-coded consistently with the horizontal cuts in the main panel, illustrating how the resulting mass function depends on the assumed AGN lifetime. We mark the PISN mass gap, $50 \lesssim m_1/\Msun \lesssim 130$, as predicted for BHs formed through standard single-star evolution \citep{sevn_2017,Woosley_2021} in gray.}
    \label{fig:timescales_for_discussion}
\end{figure}

%%%%%%%%%%%%%%%%%%%%%%%%%%%%%%%%%%%%%%%%%%%%%%%%%%%%%%%%%%%%%%%%%%%%%%%%%%%%%%%%%%%
%%%%%%%%%%%%%%%%%%%%%%%%%%%%%%%%%%%%%%%%%%%%%%%%%%%%%%%%%%%%%%%%%%%%%%%%%%%%%%%%%%%
%%%%%%%%%%%%%%%%%%%%%%%%%%%%%%%%%%%%%%%%%%%%%%%%%%%%%%%%%%%%%%%%%%%%%%%%%%%%%%%%%%%

\subsection{Robustness of the model}
\label{sec:discussion_physical_model}

%%%%%%%%%%%%%%%%%%%%%%%%%%%%%%%%%%%%%%%%%%%%%%%%%%%%%%%%%%%%%%%%%%%%%%%%%%%%%%%%%%%
%%%%%%%%%%%%%%%%%%%%%%%%%%%%%%%% FIGURES %%%%%%%%%%%%%%%%%%%%%%%%%%%%%%%%%%%%%%%%%%
%%%%%%%%%%%%%%%%%%%%%%%%%%%%%%%%%%%%%%%%%%%%%%%%%%%%%%%%%%%%%%%%%%%%%%%%%%%%%%%%%%%

\begin{figure*}
    \centering
    \includegraphics[width=0.9\linewidth]{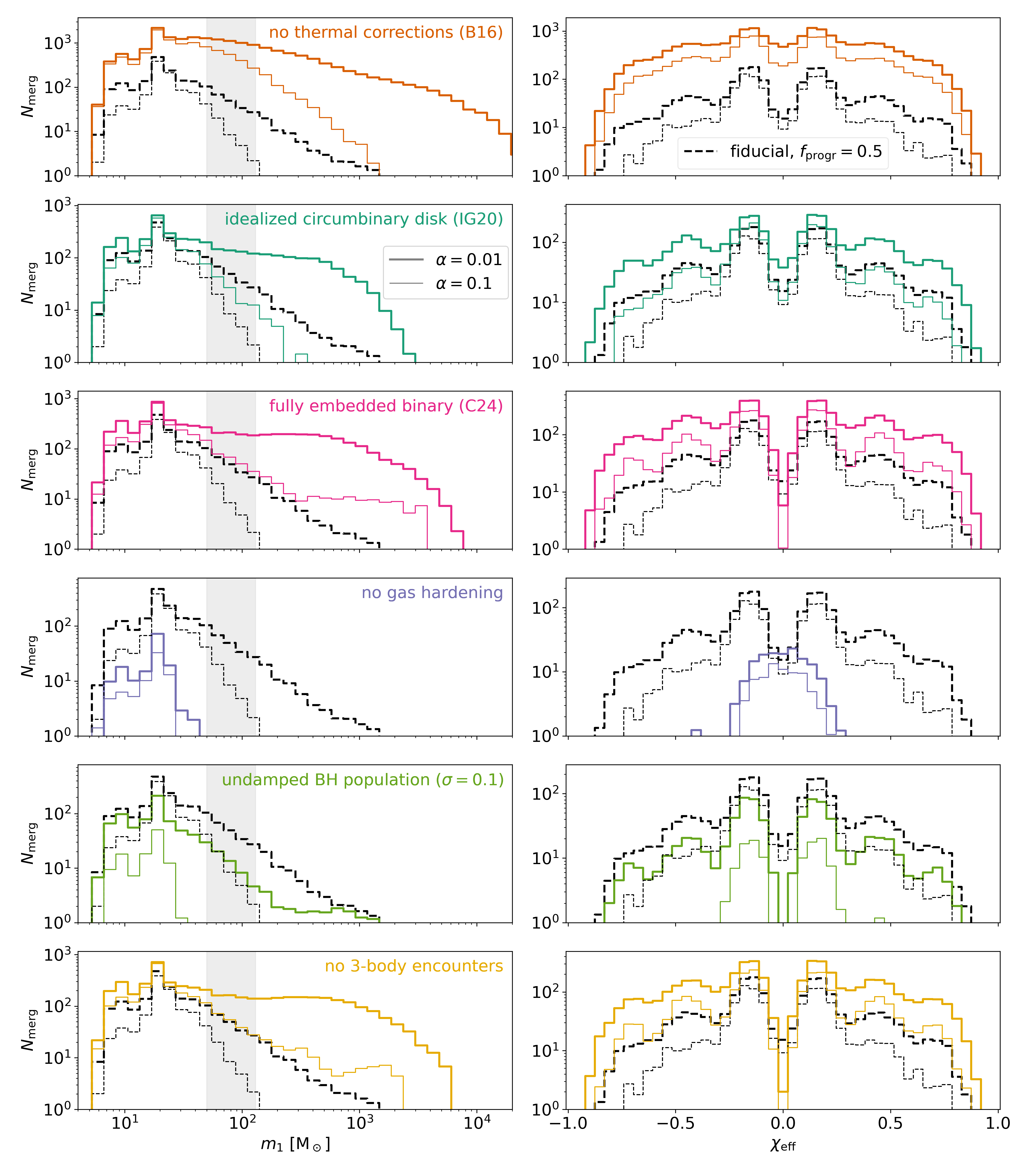}
    \caption{Each row shows a variation of the fiducial model ($f_{\rm progr}=0.5$), where a single physical ingredient is modified, as labeled in each panel: no thermal corrections \citep{Bellovary_2016}, idealized circumbinary disk \citep{Ishibashi_2020}, fully embedded binaries \citep{Calcino_2024}, no gas hardening, undamped BH population (\citealt{Trani_2024}, $\sigma=0.1$), and no three-body encounters. The left column displays the distribution of primary masses $m_1$, while the right column shows the effective spin $\chi_{\rm eff}$. We mark the PISN mass gap, $50 \lesssim m_1/\Msun \lesssim 130$, as predicted for BHs formed through standard single-star evolution \citep{sevn_2017,Woosley_2021} in gray. Colored histograms represent the modified models, while black dashed curves show the fiducial case for reference. Thick and thin lines correspond to different disk viscosities, $\alpha=0.01$ and $\alpha=0.1$, respectively.}
    \label{fig:weighted_mixture_Discussion}
\end{figure*}

\begin{figure}
    \centering
    \includegraphics[width=0.9\linewidth]{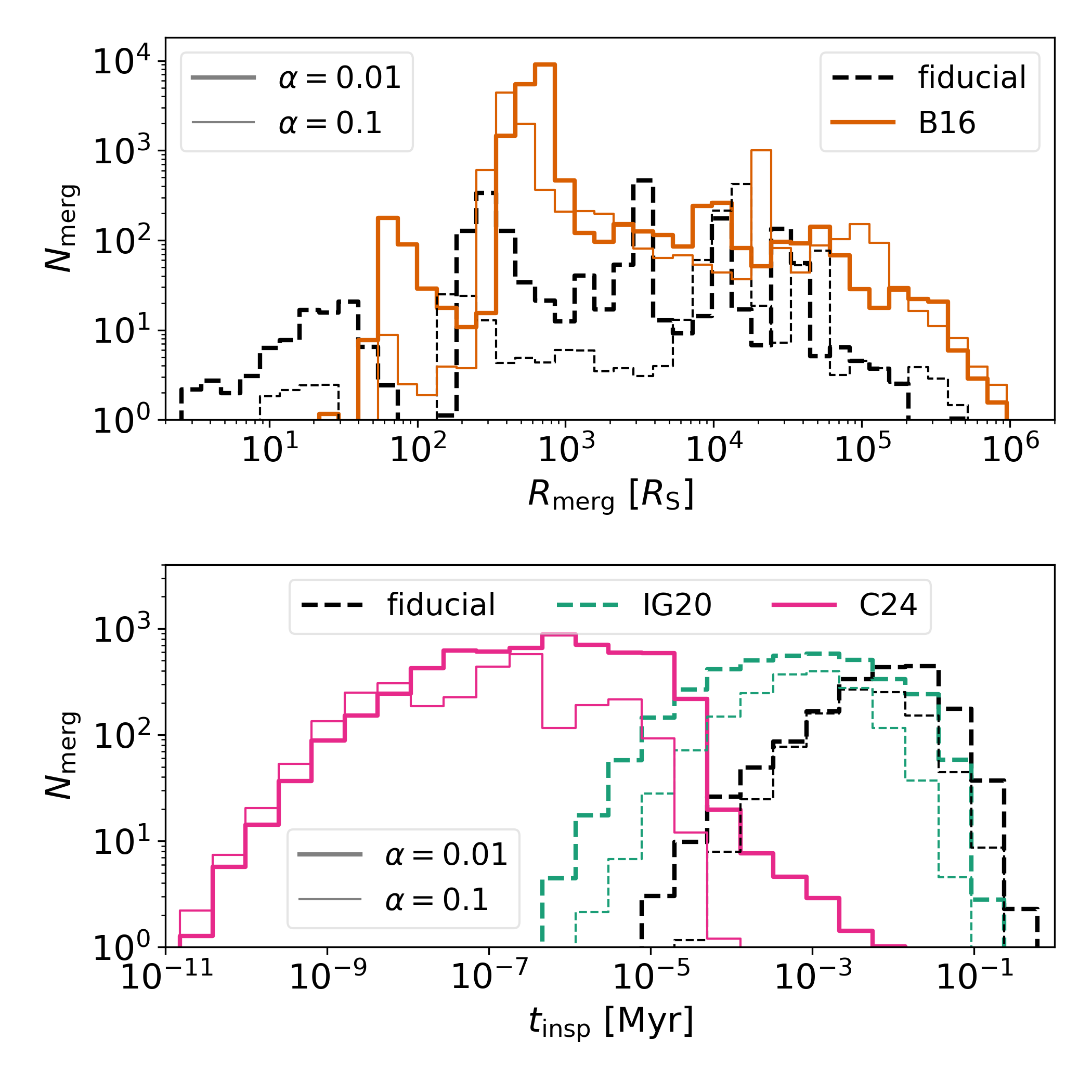}
    \caption{Top panel: distribution of merger radii $R_{\rm merg}$ in units of Schwarzschild radius, $R_{\rm S}$, for the model without thermal corrections \citep{Bellovary_2016}. Bottom panel: distribution of inspiral times, $t_{\rm insp}$, for the simplified circumbinary disk \citep{Ishibashi_2020} and the fully embedded binary prescription \citep{Calcino_2024}. %Line styles and colors follow the same convention as in \autoref{fig:weighted_mixture_Discussion}.
    }
    \label{fig:weighted_mixture_Discussion_figure_focus}
\end{figure}

%%%%%%%%%%%%%%%%%%%%%%%%%%%%%%%%%%%%%%%%%%%%%%%%%%%%%%%%%%%%%%%%%%%%%%%%%%%%%%%%%%%
%%%%%%%%%%%%%%%%%%%%%%%%%%%%%%%%%%%%%%%%%%%%%%%
%%%%%%%%%%%%%%%%%%%%%%%%%%%%%%%%%%%%%%%%%%%%%%%

Here, we assess the robustness of our results against changes in the main physical %ingredients 
processes of the model. 
In particular, we vary the prescriptions for migration torques, gas-driven binary hardening, and binary--single interactions. 
The comparison is shown in \autoref{fig:weighted_mixture_Discussion}, where we report the distributions of primary mass and effective spin for the local AGN-assisted BBH merger population, constructed as described in Section~\ref{sec:cosmo_methods}, for each model variation. 
Overall, the qualitative outcome of the model is robust: AGN disks can produce repeated mergers and populate the high-mass tail of the BBH distribution, except when gas hardening is neglected entirely. 
However, the efficiency of hierarchical growth, and therefore the extension of the high-mass tail, depends sensitively on how the disk drives radial migration and binary hardening, as well as on the properties of binary--single encounters.

\subsubsection{Different models for migration torques}
\label{sec:discussion_migration_torques}

We first compare our fiducial migration prescription with a model in which thermal corrections to the torque are neglected \citep[][hereafter B16]{Bellovary_2016}. 
We follow the implementation described in \citet[][see their Appendix B]{Vaccaro_2025}, which also includes a prescription for partial gap opening as in \citet{Kanagawa_2018}.

As shown in \autoref{fig:weighted_mixture_Discussion}, the model without thermal corrections produces a substantially larger number of mergers and a more extended high-mass tail than the fiducial model. 
A useful way to interpret this behavior is through the distribution of merger radii, shown in \autoref{fig:weighted_mixture_Discussion_figure_focus}. 
In the B16-like case, mergers preferentially occur at $300\,R_{\rm S} \lesssim R_{\rm merg} \lesssim 10^3\,R_{\rm S}$, corresponding to the location of a migration trap for a significant fraction of the contributing systems, in particular for $\log(\MSMBH/\Msun)>6.5$ ($5.8$) in the $\alpha=0.01$ ($0.1$) case \citep[see][Appendix B]{Vaccaro_2025}. 
%In contrast, the fiducial torque prescription \citep{Grishin_2024, Masset_2017} typically produces multiple traps, including outer traps at $R\simeq 10^4\,R_{\rm S}$. 
%The contribution of each individual trap is therefore less dominant in setting the overall merger-radius distribution. %, so that merger locations are less sharply peaked than in the B16 case and are instead distributed over a broader range of disk radii, extending down to $R_{\rm merg}\simeq 10\,R_{\rm S}$.
%The key difference is therefore whether merger remnants accumulate preferentially at a single migration trap, as in the B16 case, which can act as a recurrent merger site. The enhanced concentration of mergers near a trap implies that remnants can remain in a region where they efficiently pair up again, producing more and higher-generation BBH mergers. 

The key difference with respect with the fiducial torque prescription \citep{Grishin_2024, Masset_2017} is how efficiently migration concentrates BHs at specific disk radii. In the B16-like model, many systems are driven toward the same inner migration trap. This single trap then acts as a preferred ``meeting point'' for BHs: binaries form and merge there, and the merger products are also likely to remain in, or return to, the same region. As a result, the same location can be reused over multiple merger generations. Moreover, as shown in \autoref{fig:pAGN_profiles}, $R \lesssim 10^3\,R_{\rm S}$ corresponds to the location of the disk where the gas is densest, shortening the timescales of dynamical processes such as capture (see \autoref{app:gas_capture}), migration \citep{Vaccaro_2025}, pairing (eq.~\ref{eq:t_pair}), and binary-single encounters (eq.~\ref{eq:t_enc}) . This leads to a larger total number of mergers and to a more extended high-mass tail in the B16-like model. 

By contrast, in the fiducial torque prescription \citep{Grishin_2024, Masset_2017}, BBH mergers are distributed more uniformly over a broader range of disk radii, meaning that the contribution of each individual merger location is less dominant in setting the overall $R_{\rm merg}$ distribution. The fiducial model therefore does not select a single, dense inner region as a recurrent merger site. Instead, merger remnants are spread across different disk environments, including outer traps where the gas surface density is lower and the relevant timescales for capture, pairing, migration, and binary-single encounters are longer. As a result, the merger-product population undergoes less efficient dynamical processing, suppressing the production of high-generation BBH mergers. %Overall, including thermal corrections to the Type~I migration torques as in \citet{Grishin_2024} makes hierarchical growth less efficient.}

\subsubsection{Different models for gas hardening}
\label{sec:discussion_gas_hardening}

We next investigate the impact of the gas-hardening prescription. 
Our fiducial model adopts the \citet[][hereafter \citetalias{Ishibashi_2024}]{Ishibashi_2024} prescription, and we compare it to a simplified circumbinary-disk model following \citet[][hereafter \citetalias{Ishibashi_2020}]{Ishibashi_2020}, to the fully embedded binary prescription of \citet[][hereafter \citetalias{Calcino_2024}]{Calcino_2024}, and finally to a limiting case in which gas hardening is switched off.

These models correspond to different assumptions about how a BBH exchanges angular momentum with the surrounding gas. 
Both \citetalias{Ishibashi_2020} and \citetalias{Ishibashi_2024} describe the interaction between a BBH and its circumbinary disk. 
In \citetalias{Ishibashi_2020}, mass accretion onto the binary is neglected, and the binary always loses angular momentum to the disk and hardens. \citetalias{Ishibashi_2024} extends this framework by explicitly including gas accretion onto the binary. As a result, the sign of the orbital evolution is no longer fixed: depending on the thickness of the circumbinary disk, the binary can either harden or soften. 
The \citetalias{Calcino_2024} prescription describes a different physical regime, in which the binary is fully immersed in the surrounding gas, rather than clearing a circumbinary cavity. 
Finally, the model with no gas hardening provides a limiting case where BBHs can shrink only through three-body encounters and GW emission. 
This comparison therefore brackets possible gas-driven effects. 

The comparison in \autoref{fig:weighted_mixture_Discussion} shows that the \citetalias{Ishibashi_2020} prescription produces a higher merger efficiency and a more extended high-mass tail than our fiducial \citetalias{Ishibashi_2024} model. 
This difference follows directly from the explicit inclusion of gas inflow onto the binary in \citetalias{Ishibashi_2024}: \citetalias{Ishibashi_2020} effectively enforces orbital contraction, whereas \citetalias{Ishibashi_2024} allows the BBHs to either harden or soften depending on the properties of the circumbinary disk.

The embedded-binary case leads to an even more pronounced increase in the merger efficiency. We model this case by adopting the $\dot a$ and $\dot e$ values calibrated from the hydrodynamical simulations of \citetalias{Calcino_2024} directly from their Figure~10. 
When implemented at face value, these rates yield extremely short inspiral times, allowing BBHs to merge rapidly after formation. Gas-driven inspiral times reach values as short as $\sim 10^{-10}\,\rm Myr$, i.e. less than an hour, as shown in \autoref{fig:weighted_mixture_Discussion_figure_focus}.
While this illustrates the potentially strong impact of gas torques in the embedded regime, it also suggests that applying the \citetalias{Calcino_2024} prescription throughout the entire BBH evolutionary history may be too simplified. 
In particular, the adopted $\dot a$ and $\dot e$ values are time-averaged rates calibrated for a specific orbital configuration, but are here applied continuously as the BBH separation evolves. 
The resulting merger efficiencies should therefore be interpreted as an exploratory limiting case, rather than as predictions of a fully self-consistent alternative model.

The no gas hardening case provides a complementary limiting case. 
In this model, BBH evolution is driven only by dynamical encounters and GW emission. 
As expected, removing gas hardening strongly suppresses the merger efficiency. 
The resulting population does not include any systems with primary masses above $40\Msun$, and loses any signatures related to angular momentum alignment in the effective spin distribution \citep[see also][]{Vaccaro_2023}, confirming that gas hardening is one of the main ingredients that controls how efficiently AGN disks can build up massive remnants through hierarchical growth.

A further possibility, not explicitly captured by our models, is that gas hardening may stall at intermediate separations. Gaseous dynamical friction is expected to become less efficient when the binary motion relative to the gas becomes highly supersonic (Mach number $\mathcal{M}\gtrsim5$; \citealt{Ostriker_1999, Kim_2008}). Since the orbital velocity increases as the binary shrinks, this regime may be reached before GW emission becomes efficient, leaving a range of separations in which neither mechanism can rapidly drive the binary towards merger. After stalling, the binary would evolve similarly to our no-gas-hardening case and may require a three-body encounter to reach the GW-dominated regime, possibly retaining higher eccentricities in the GW band.

\subsubsection{Different models for three-body encounters}
\label{sec:discussion_three_body}

Finally, we investigate the role of binary--single encounters. 
In our fiducial model, the tertiary BH population is assumed to be dynamically cold, with eccentricities and inclinations efficiently damped by the gas, corresponding to $\sigma_{\rm BH}<10^{-4}$ (see \autoref{app:binary_evolution}). 
We compare this model to two limiting cases: one in which three-body encounters are neglected, and one in which the embedded BH population is dynamically warmer, with $\sigma_{\rm BH}=0.1$.

As shown in \autoref{fig:weighted_mixture_Discussion}, neglecting three-body encounters increases the merger efficiency and produces a more extended high-mass tail. 
This indicates that binary--single encounters interrupt the otherwise smooth gas-driven inspiral of BBHs. Indeed, in our framework, the total time to merger includes both the gas-driven inspiral time after the latest encounter and the cumulative time spent waiting for encounters, as in eq.~\ref{eq:inspiral_timescale}. This additional time budget may reduce the probability that a merger occurs within the finite AGN lifetime relative to a model in which three-body encounters are neglected.

%encounters introduce an additional delay since the effective inspiral time includes both the gas-driven inspiral time and the cumulative time spent between encounters, as in eq.~\ref{eq:inspiral_timescale}.
%This additional delay reduces the probability that a binary merges within the finite AGN lifetime with respect to a framework where three-body encounters are neglected.

Conversely, in the dynamically warmer model with $\sigma_{\rm BH}=0.1$, hierarchical growth is suppressed. 
With this velocity dispersion, binary--single encounters are more likely to break up, or ionize, BBHs than in the dynamically cold case ($\sigma_{\rm BH}<10^{-4}$), where encounters are dominated by Keplerian shear and the binaries are effectively hard \citep{Trani_2024}. 
We discuss this behavior further in \autoref{app:3bb}. 

Overall, a warmer population of BHs ($\sigma_{\rm BH}=0.1$) reduces the merger efficiency and limits the build-up of higher-generation remnants.

\subsection{Limitations of the model}

Our semi-analytical framework relies on several simplifying assumptions that should be kept in mind when interpreting the results. Here, we adopt a simplified framework that isolates the role of hierarchical mergers in AGN disks. We do not include in-situ star formation or stellar evolution during the AGN episode, and we neglect mass growth through gas accretion onto BHs. We do not model the dynamical evolution of the surrounding nuclear star cluster (NSC) nor the replenishment of captured BHs by two-body relaxation in the NSC. As a result, %the BH population available for migration and binary formation is fixed at the beginning of the AGN episode, and 
the only mechanism driving BH mass growth in our model is repeated mergers.

Our fiducial AGN disk model \citepalias{SG}  
assumes an unspecified heating source and does not account for gas consumption even in gravitationally unstable regions. 
Moreover, it neglects magnetic fields, which could prevent spin alignment \citep{Dhruv_2025}, and assumes a fixed viscosity parameter, $\alpha$, throughout the entire disk. We also neglected time evolution and vertical stratification in the AGN disk. We assumed all BHs to migrate strictly in the midplane after orbital damping, thereby omitting any effect related to vertical motion. We also neglect turbulence, which may prevent the full circularization and alignment of BH orbits \citep{Trani_2025} and introduce stochastic fluctuations on top of the mean Type I migration torque, causing BHs to wander radially and broadening the distribution of their encounter locations \citep{Wu_2024}. If sufficiently strong, turbulence could reduce the efficiency of repeated mergers at convergence regions, suppressing the formation of the highest-generation BHs and producing more broadly distributed merger radii.

We rely on quasar proximity-zone measurements \citep{Khrykin_quasar_lifetime} to estimate the lifetime of AGN disks, $\tau$. These measurements constrain the lifetime of the radiatively bright phase, but they do not necessarily correspond to the full lifetime of the gaseous disk. In particular, the disk may persist, at least partially, during phases in which the AGN is weakly accreting or no longer observable as a luminous quasar.
In addition, we modeled each AGN episode as a single continuous phase of duration $\tau$ and neglected any previous AGN activity. However, AGNs may undergo multiple bursts of activity separated by quiescent phases, potentially allowing BHs to survive between episodes and continue their migration and merger history during subsequent active phases. Accounting for repeated AGN episodes could therefore increase the probability of hierarchical mergers and modify the predicted merger delay time distribution.

Our pair-up prescription remains qualitative. In our current model, we adopted a timescale requirement (eq.~\ref{eq:t_pair}, \citealt{Qian_2024}) to evaluate whether two nearby BHs will form a bound binary. However, hydrodynamical simulations show that both the efficiency of pair-up and the properties of the resulting binary%, such as its semi-major axis, eccentricity, and orbital orientation, 
depend sensitively on the orbital configuration of the interacting BHs, such as their relative inclination, and on the local gas properties \citep[e.g.][]{Qian_2024, Whitehead_2024}. Therefore, our treatment of $f_{\rm progr}$ is  simplified, as the fraction of prograde and retrograde binaries should depend on the disk location and on the details of the interaction. This may affect the predicted effective and precession spin distributions of the BBH population. 
Furthermore, we do not consider dynamical BBH formation through three-body scatterings, which is another viable formation channel in AGN environments \citep{Tagawa_2020}. Such interactions may contribute significantly in dense regions of the disk and could produce BBHs with different orbital and spin properties than those predicted by our migration-driven pair-up model.

Finally, our model does not include resonant or secular dynamical effects, such as mean-motion resonances, which could alter or even halt migration \citep{EpsteinMartin_2025, Moncreiff_2025}, or von~Zeipel-Kozai-Lidov oscillations, which may become important in the innermost regions of the disk by exciting the eccentricity and inclination of BBHs \citep{Su_2025}. However, as discussed in \autoref{app:timescales}, the von~Zeipel-Kozai-Lidov timescale is generally longer than the inspiral timescale of our binaries, suggesting that these effects are unlikely to play a major role in most of our systems.

Addressing these limitations will require a combination of more realistic AGN disk models, improved treatments of binary formation and remnant properties, and dedicated hydrodynamical and $N$-body simulations, which we defer to future work.

\section{Summary}
\label{sec:conclusions}

We presented an updated semi-analytical framework to model the formation and hierarchical growth of BBHs in AGN disks. 
Starting from populations of stellar-origin BHs embedded in a gaseous disk, we followed their capture, migration, pair-up, gas-driven hardening, dynamical encounters, and eventual merger. 
We systematically explored how the resulting BBH population depends on the main properties of the AGN environment, namely the SMBH mass, $\MSMBH$, its accretion rate, $\fEdd$, and the disk viscosity parameter, $\alpha$. 
We then constructed an intrinsic local BBH merger population by weighting  individual simulations according to the observed distribution of low-redshift AGN properties.

Our main results can be summarized as follows.

\begin{itemize}
    \item[$\bullet$] Hierarchical growth in AGN disks is highly sensitive to the properties of the disk environment. 
    In particular, lower-viscosity \citetalias{SG} disks, with $\alpha=0.01$, allow for more sustained hierarchical growth than disks with $\alpha=0.1$, reaching higher generations and producing more massive remnants.

    \item[$\bullet$] Gas surface density and angular-momentum transport are key regulators of hierarchical growth. Dense, low-viscosity disks promote efficient capture, migration, pairing, and hardening, enabling more mergers and longer hierarchical merger chains within the AGN lifetime. As shown in \autoref{app:TQM}, this trend extends beyond locally viscous $\alpha$-disk models: the rapid global inflow and low surface density of \citet{TQM} disks produce fewer, later mergers and strongly suppress the formation of a high-mass BBH merger population.

    %%\item \st{The AGN channel does not have a single characteristic outcome across the $(\MSMBH,\fEdd)$ parameter space.     Instead, we find a broad range of behaviors, from nearly inactive systems to environments capable of efficient repeated mergers. The merger efficiency generally increases with $\fEdd$, but the dependence on $\MSMBH$ is non-monotonic. In particular, large SMBH masses alone do not guarantee efficient BBH assembly: in some high-$\MSMBH$ configurations, mergers are strongly suppressed or absent. }
    %This highlights that BBH production in AGN disks is controlled by the underlying disk structure, rather than by any single global AGN parameter.

    \item[$\bullet$] The local BBH merger population from the AGN channel extends from stellar-mass BBHs to a high-mass tail produced by hierarchical mergers. 
    This tail can reach $m_1 \gtrsim 10^3\,\Msun$ in the most favorable configurations, and it is more prominent for low-viscosity disks ($\alpha=0.01$). 
    
    \item[$\bullet$] The same hierarchical process also drives the population toward low mass ratios, especially for massive primaries. This occurs because repeated mergers preferentially increase the mass of the primary, while the secondary mass saturates below the PISN mass gap. 
    In our hierarchical pairing prescription, high-generation merger products can merge with lower-generation companions, producing increasingly unequal-mass systems. 
    As a result, the AGN channel naturally populates the low-$q$ region of the BBH parameter space.

    \item[$\bullet$] The spin distributions retain a strong imprint of the AGN disk geometry. 
    The effective spin distribution depends sensitively on the fraction of BBHs that are prograde at formation, $f_{\rm progr}$: fully prograde populations are biased toward positive $\chi_{\rm eff}$, fully retrograde populations toward negative $\chi_{\rm eff}$, and mixed populations develop a bimodal structure. 
    
    \item[$\bullet$] We find a correlation between $m_1$ and $|\chi_{\rm eff}|$, reflecting the coherent build-up of angular momentum through repeated mergers in the disk. 
    More massive primaries are typically the products of longer merger chains, and therefore retain memory of the preferred angular-momentum direction set by the disk. 
    This leads to larger values of $|\chi_{\rm eff}|$ at high masses.%, with the sign of $\chi_{\rm eff}$ influenced by the initial prograde or retrograde configuration. 
    %This trend is a characteristic outcome of hierarchical growth in a flattened, gas-rich environment. 

    \item[$\bullet$] When compared with massive BBH events in GWTC-4, GW190521 and GW231123, our local AGN population can reach broadly compatible regions of parameter space, but these events lie close to the edge of the predicted distributions.The primary mass and effective spin inferred for GW190521 (GW231123) can be reproduced at rates of one in $60$ ($200$) mergers in our lower-viscosity models ($\alpha=0.01$). 
    The main difficulty is instead producing systems that also contain secondary with a large enough mass. 
    As a representative measure, in the $(m_1,m_2)$ plane GW190521-like systems occur at the level of roughly one in $500$ AGN mergers, while GW231123-like systems occur at the level of roughly one in $3000$ mergers in the $\alpha=0.01$ models. 
    %In particular, GW231123 falls in a region where the model tends to underestimate its secondary mass. 
    This suggests that such events are possible within the AGN channel, but are expected to be relatively rare in the present implementation. 

    \item[$\bullet$] The AGN lifetime is a key regulator of hierarchical growth.     Longer accretion episodes allow merger chains to proceed to higher generations and produce a more extended high-mass tail. In our model, the formation of IMBHs with $m_1 \gtrsim 100\,\Msun$ ($1000\,\Msun$) AGN episodes longer than  $\tau_{\rm max} \gtrsim 1$ Myr ($10$ Myr). 

\end{itemize}

Overall, our results show that AGN disks provide a viable environment for the formation of massive and hierarchical BBH mergers. This BBH population is characterized by increasingly unequal mass ratios at high primary mass, and an effective-spin distribution that depends strongly on first-generation spin magnitudes and on the fraction of binaries born in prograde or retrograde configurations. %but that their predictions are strongly shaped by uncertain AGN disk physics. 
Here, we focused on the local BBH population and on the dependence of AGN-assisted mergers on the underlying disk physics. In a companion paper, we will extend this framework to a cosmological population of AGNs, in order to explore the redshift evolution of the AGN disk scenario and its implications for third-generation GW detectors.

%Overall, our results show that AGN disks provide a viable environment for the formation of massive and hierarchical BBH mergers, but that their predictions are strongly shaped by uncertain disk and binary-evolution physics. The most distinctive signatures of this channel are not a single feature, but a combination of properties: a high-mass hierarchical tail, increasingly unequal mass ratios at high primary mass, correlations between mass and effective spin, a dependence of $\chi_{\rm eff}$ on the prograde or retrograde configuration of the binary, and a subpopulation of eccentric mergers. At the same time, the rarity of the most massive systems and the sensitivity to the gas-hardening, migration, and encounter prescriptions show that robust predictions require further calibration against dedicated hydrodynamical and few-body simulations.

%The present work focuses on the local BBH population and on the dependence of AGN-assisted mergers on the underlying disk physics. In a companion paper, we will extend this framework to a cosmological population of AGNs and SMBHs, connecting the merger yields derived here to the redshift evolution of the AGN population and assessing the implications for third-generation GW detectors such as the Einstein Telescope and Cosmic Explorer.

\section*{Data availability}
The code used in this study is available via Gitlab at \href{https://gitlab.com/mariapaola.vaccaro/fastcluster_agn#}{gitlab.com/mariapaola.vaccaro/fastcluster$\_$agn}.
Supplementary material and data are available via Zenodo at \href{https://zenodo.org/records/22078875}{zenodo.org/records/22078875}.

%%%%%%%%%%%%%%%%%%%%%%%%%%%%%%%%%%%%%%%%%%%%%%%%%%%%%%%%%%%%%%
\begin{acknowledgements}
We thank the anonymous referee for constructive comments. We thank Lumen Boco, Dominika Wylezalek, Ralf Klessen, Laura Sberna, Jupiter Stevenson, Manuel Arca Sedda, Biswajit Banerjee, Josh Calcino, Ilya Khrykin, Debora Sijiacki, Roberto Maiolino, Om Sharan Salafia, Zoltan Haiman, Mark Avara, Samson Leong, Yannic Pietschke, and Douglas Lin for useful discussions. 

MPV, MM, and BL acknowledge financial support from the European Research Council for the ERC Consolidator grant DEMOBLACK, under contract no. 770017. They also acknowledge financial support from the Deutsche Forschungsgemeinschaft (DFG, German Research Foundation) under Germany's Excellence Strategy EXC 2181-390900948 (the Heidelberg STRUCTURES Excellence Cluster). MPV acknowledges support from the Scientific Software Center (SSC) at Heidelberg University through the SSC Fellows program, and is especially grateful to Tuyen Le and 
Dmitrii Kapitan.
The authors acknowledge support by the state of Baden-W\"urttemberg through bwHPC and the German Research Foundation (DFG) through grants INST 35/1597-1 FUGG and INST 35/1503-1 FUGG. 

Our simulations are based on \href{https://gitlab.com/micmap/fastcluster_open}{\fastcluster{}} \citep{fastcluster2021}. We also made use of \href{https://gitlab.com/aatrani/pagn}{\pagn{}} \citep{Gangardt_2024}, \href{https://gitlab.com/sevncodes/sevn}{\sevn{}} \citep{sevn_2023}, \tsunami{} \citep{tsunami}, 
{\fontfamily{lmtt}\selectfont NumPy} \citep{Harris20}, {\fontfamily{lmtt}\selectfont SciPy} \citep{SciPy2020}, {\fontfamily{lmtt}\selectfont Pandas} \citep{Pandas2024}. Plots were produced using {\fontfamily{lmtt}\selectfont Matplotlib} \citep{Hunter2007}.
\end{acknowledgements}

\bibliographystyle{aa} % style aa.bst
\bibliography{bibliography.bib}

\appendix
\section{AGN disk structure}\label{app:pagn}
\begin{figure}
    \centering
    \includegraphics[width=\linewidth]{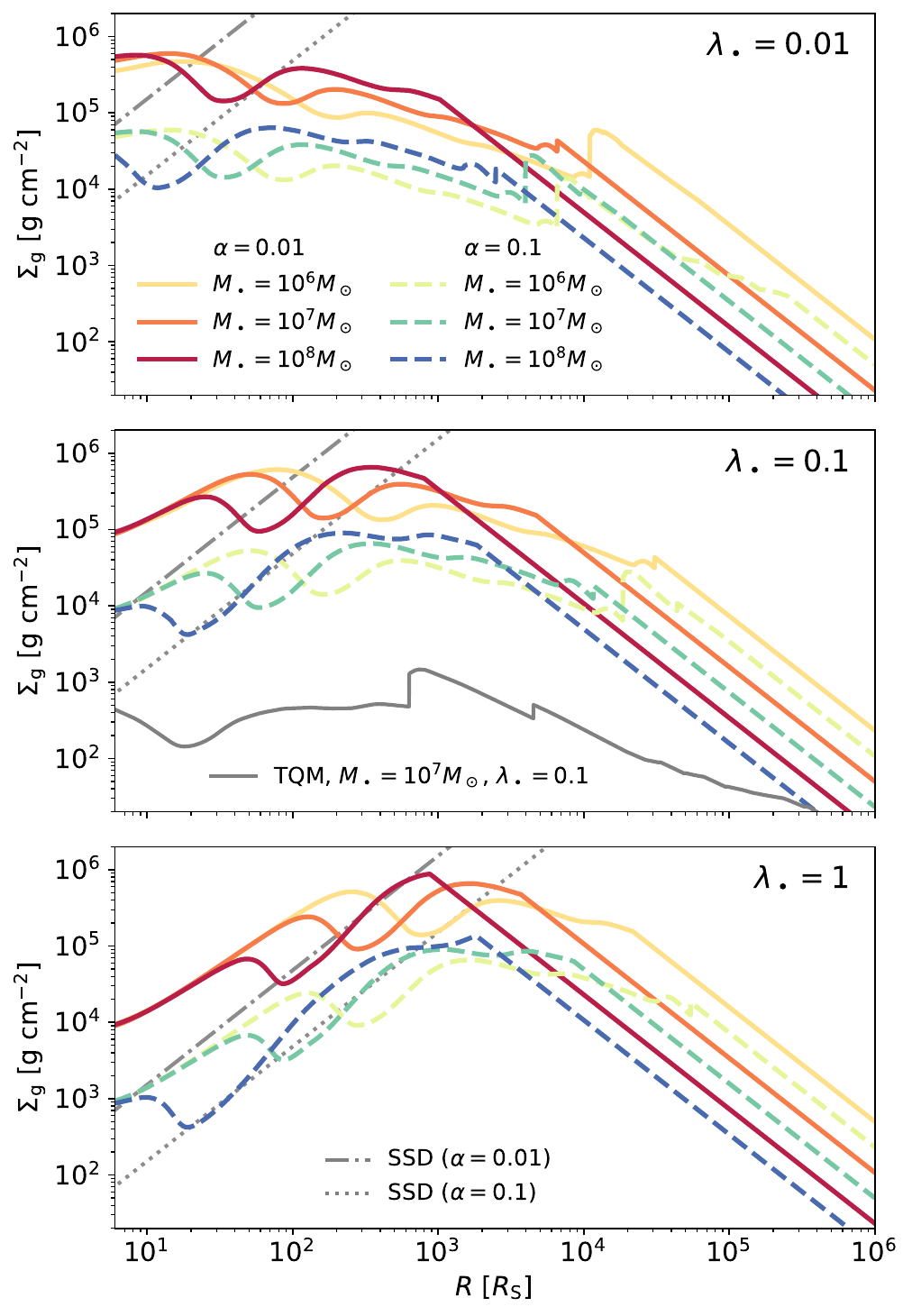}
    \caption{Gas surface-density profiles, $\Sigma_{\rm g}$, as a function of radius for the \citet{SG} AGN disk model \citep[computed with \pagn{},][]{Gangardt_2024}. Different panels correspond to different SMBH Eddington ratios, $\fEdd=0.01, 0.1,$ and $1$ from top to bottom. Colored curves show disk profiles for $\MSMBH=10^6, 10^7,$ and $10^8\,\Msun$. Color intensity encodes $\MSMBH$, while color family and line style distinguish the viscosity parameter: $\alpha=0.01$ in red solid lines and $\alpha=0.1$ in blue dashed lines. Grey lines show a \citet{TQM} disk profile for $\MSMBH = 10^7 \Msun$ and $\fEdd=0.1$ \citep[computed with \pagn{},][]{Gangardt_2024}, and \citet{SSD} power-law scalings \citep[as in][]{Kocsis_2011}.
}
    \label{fig:pAGN_profiles}
\end{figure}

We model the radial structure of an AGN disk with the steady-state \citet{SG} disk model, computed with the \pagn{} module \citep{Gangardt_2024} for each combination of $\MSMBH$, $\fEdd$, and $\alpha$ considered in our parameter grid. 
\autoref{fig:pAGN_profiles} shows representative gas surface-density profiles, $\Sigma_{\rm g}$, as a function of radius. Radii are in units of the Schwarzschild radius, $R_{\rm S}$, so that differences between disks with different SMBH masses can be compared on the same dimensionless radial scale. 
At fixed $\MSMBH$ and $\fEdd$, disks with $\alpha=0.01$ are systematically denser than the corresponding disks with $\alpha=0.1$, typically by about an order of magnitude. 
For comparison, we also show representative Shakura--Sunyaev $\alpha$-disk scalings \citep{SSD}, computed using the \citet[][eqs.~32--33]{Kocsis_2011} treatment, in which quantities such as the gas surface density, $\Sigma_{\rm g}$, and aspect ratio, $h$, are modeled as power laws. 

\section{Comparison with the TQM disk model}
\label{app:TQM}

\begin{figure}
    \centering
    \includegraphics[width=\linewidth]{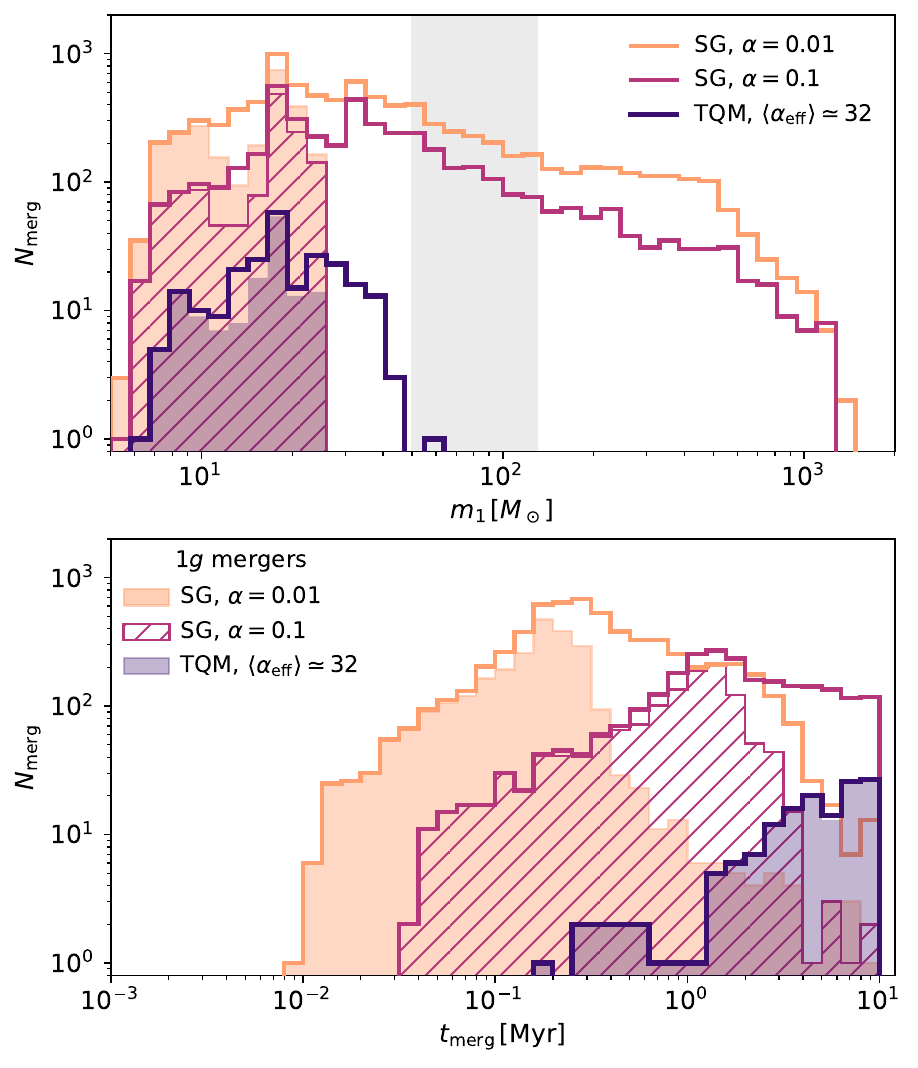}
    \caption{Distributions of the primary BH mass, $m_1$, %effective spin, $\chi_{\rm eff}$, 
    and merger timescale, $t_{\rm merg}$, for simulations with $\MSMBH=10^7\,M_\odot$, $\fEdd=0.1$, and a fixed AGN lifetime of $10\,\mathrm{Myr}$. We use the \citetalias{Calcino_2024} prescription for BBH evolution, and assume $f_{\rm progr}=1$. We compare the \citetalias{SG} model with $\alpha=0.01$ (light orange) and $0.1$ (pink) to the \citetalias{TQM} model (dark purple). Solid outlines show the distributions of all mergers, while filled or hatched histograms show the first-generation subset. We mark the PISN mass gap, $50 \lesssim m_1/\Msun \lesssim 130$, as predicted for BHs formed through standard single-star evolution \citep{sevn_2017,Woosley_2021} in gray.}
    \label{fig:TQM_SG_comparison}
\end{figure}

As a test of the dependence of our results on the adopted disk model, we repeated our calculation using the \citet{TQM} (hereafter \citetalias{TQM}) model and compared it with our fiducial \citetalias{SG} model. Unlike the \citetalias{SG} model, in which angular-momentum transport is described through a local viscosity parameter, $\alpha$, the \citetalias{TQM} model assumes rapid inflow driven by global gravitational torques, with radial velocity $v_R=m_{\rm T}\,c_{\rm s}$. We set $m_{\rm T}=0.1$, corresponding to a rapid but subsonic inflow, as assumed by \citetalias{TQM}. We adopt the fiducial \citetalias{TQM} parameters for stellar radiative and supernova feedback, $\epsilon_{\rm T}=10^{-3}$ and $\xi=1$ respectively. We set the disk boundaries to $R_{\rm in}=3\,R_{\rm S}$ and $R_{\rm out}=10^7\, R_{\rm S}$, and use the combined opacity prescription implemented in \pagn{} \citep{Gangardt_2024}.
In the outer \citetalias{TQM} disk, ongoing star formation maintains the disk near marginal gravitational stability while consuming part of its gas reservoir. Consequently, the accretion rate varies with radius. We determine $\dot{M}_{\rm out}=\dot{M}(R_{\rm out})$ iteratively by requiring that the rate reaching the inner boundary satisfies $\dot{M}(R_{\rm in})=\fEdd \dot{M}_{\rm Edd}$. Because each trial value of $\dot{M}_{\rm out}$ requires a new radial integration, this calculation is slower than the direct \citetalias{SG} solution. We therefore use the \citetalias{TQM} model as a representative comparison for $\MSMBH=10^7\Msun$ and $\fEdd=0.1$, rather than repeating the calculation over the full parameter grid.

As shown in \autoref{fig:pAGN_profiles}, the resulting \citetalias{TQM} disk has a substantially lower gas surface density than a corresponding \citetalias{SG} disk with the same $\MSMBH$ and $\fEdd$. This difference arises from gas consumption through star formation and from the rapid radial inflow assumed by the \citetalias{TQM} model, $m_T=0.1$. For interpretative purposes, the transport efficiency of the \citetalias{TQM} disk can be expressed in a form analogous to the viscosity parameter of an $\alpha$-disk by equating the \citetalias{TQM} mass-inflow rate, $\dot{M}=2\pi R\Sigma_{\rm g}m_{\rm T} c_{\rm s}$, with the standard steady-state relation for a Keplerian $\alpha$ disk, $\dot{M}=3\pi\alpha_{\rm eff}\Sigma_{\rm g}c_{\rm s}H$. This gives an effective viscosity parameter %\footnote{This quantity should be interpreted only as the local value of $\alpha$ that would reproduce the same radial mass flux under the assumptions of a steady Keplerian $\alpha$ disk. It does not imply that angular momentum is transported by a local viscosity in the \citetalias{TQM} model, nor does it provide a unique mapping between the two disk prescriptions.}
$\alpha_{\rm eff}=2m_{\rm T}/3h$, where $h=H/R$. For the disk considered here, $\alpha_{\rm eff}(R)$ ranges from approximately $1$ to $100$, far above the values adopted in \citetalias{SG} models. This quantity is only a diagnostic: it illustrates the efficiency of the assumed global inflow and should not be interpreted as a physical viscosity or as a unique mapping between the two disk models. We therefore retain the fiducial $m_{\rm T}=0.1$ rather than tuning it to reproduce a chosen value of $\alpha_{\rm eff}$, which would substantially alter the fueling regime assumed in the original \citetalias{TQM} model.

We adopt a fixed AGN lifetime of $10 \, \mathrm{Myr}$, as in \citet{Tagawa_2020}, to allow sufficient time for dynamical interactions in the comparatively low-density \citetalias{TQM} disk. We also use the \citetalias{Calcino_2024} prescription for gas-driven binary evolution instead of our fiducial \citetalias{Ishibashi_2024} prescription, because the latter explicitly assumes angular-momentum transport through a local $\alpha$ viscosity and is therefore not directly applicable to the \citetalias{TQM} model. For a consistent comparison, we repeated the \citetalias{SG} calculations with $\alpha=0.01$ and $0.1$ using the same AGN lifetime and binary-hardening prescription. We ran $10^4$ independent Monte Carlo realizations for each disk model.

The resulting distributions are shown in \autoref{fig:TQM_SG_comparison}. The \citetalias{SG} disk with $\alpha=0.01$ produces the largest number of mergers and the broadest hierarchical tail, extending to primary masses above $10^3\Msun$, with typical merger timescales of $t_{\rm merg}\sim10^5 \,\mathrm{yr}$. Increasing the viscosity to $\alpha=0.1$ lowers the surface density of the \citetalias{SG} disk, shifts the distribution toward $t_{\rm merg}\sim1\,\mathrm{Myr}$, and limits the extent of hierarchical growth. The \citetalias{TQM} model produces substantially fewer mergers, predominantly at $t_{\rm merg}> 1\,\mathrm{Myr}$. Although a higher-generation component remains visible in the primary-mass distribution, it only sparsely reaches the lower edge of the PISN mass gap and does not develop the extended high-mass population found in the denser \citetalias{SG} disks.

These results support a physical interpretation based primarily on the gas surface density and on the efficiency of angular-momentum transport. At a fixed accretion rate reaching the SMBH, $\dot{M}(R_{\rm in})$, a larger radial inflow velocity generally allows the same mass flux to be sustained by a smaller gas column. This is reflected in the decrease of $\Sigma_{\rm g}$ with increasing $\alpha$ or $\alpha_{\rm eff}$. The lower surface density weakens gas-driven capture, migration, binary pairing, and hardening, thereby increasing the time required for mergers to occur. Consequently, fewer binaries merge within the AGN lifetime and merger remnants have fewer opportunities to undergo repeated mergers. Thus, while $\alpha$ and $\alpha_{\rm eff}$ should not be interpreted as directly equivalent parameters, the \citetalias{TQM} model extends the same qualitative trend to its extreme: more efficient transport is associated with fewer mergers and less hierarchical growth.

\section{Stellar-origin BHs with low natal spins}\label{app:chi_zero}

\begin{figure*}
    \centering
    \includegraphics[width=\linewidth]{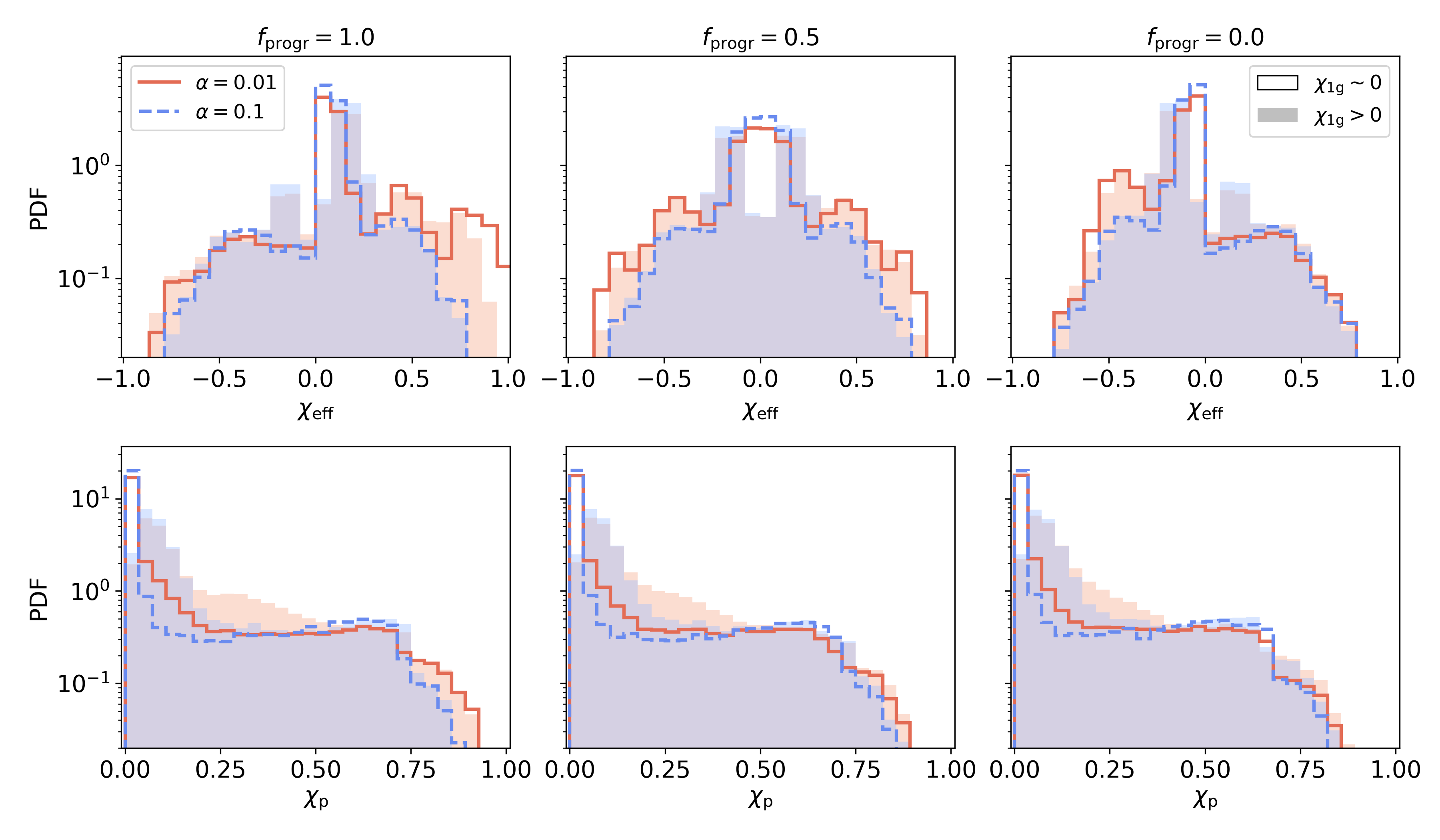}
    \caption{Probability density distributions of the effective inspiral spin, $\chi_{\rm eff}$, and the effective precession spin, $\chi_{\rm p}$, for the local AGN-disk merger population in the initially quasi-non-spinning BH scenario. 
The three columns correspond to different fractions of initially prograde BH orbits, $f_{\rm progr}=1$, $0.5$, and $0$, from left to right. 
Red and blue curves show the $\chi_{\rm eff}$ and  $\chi_{\rm p}$ distributions for $\alpha=0.01$ and $\alpha=0.1$, respectively, in the case of low natal spins ($\chi_{\rm 1g}\sim0$). The filled histograms show the corresponding distributions of our fiducial scenario ($\chi_{\rm 1g}>0$). 
}
    \label{fig:spins_chi_zero}
\end{figure*}

The spin magnitude of stellar-origin BHs remains uncertain. While current gravitational-wave observations are broadly consistent with low natal spins for the majority of stellar-mass BHs \citep[e.g.][]{GWTC4_pop}, the exact shape of the stellar-origin distribution is still poorly constrained. In our fiducial models, we therefore adopt a Maxwellian distribution for the dimensionless spin magnitudes with one-dimensional root-mean-square $\sigma_\chi=0.05$, motivated by the spin magnitudes inferred from the GWTC-4 population analysis. However, even such modest spin amplitudes can produce visible structure in the resulting $\chi_{\rm eff}$ distributions after repeated hierarchical mergers, as discussed in \autoref{sec:local_pop}.
To assess how sensitive our results are to the assumed natal-spin distribution, we also consider an initially quasi-non-spinning BH population. In this scenario, the spin magnitudes of first-generation BHs are drawn from a truncated Gaussian distribution centered on zero, $p(\chi_1) \sim \mathcal{N}(0, \sigma_\chi)$,
with $\sigma_\chi=0.05$ and truncated at $\chi\in[0,1]$. This prescription strongly suppresses the abundance of moderately spinning first-generation BHs while still allowing hierarchical mergers to build up larger spins in subsequent generations.

\autoref{fig:spins_chi_zero} shows the resulting distributions of the effective inspiral spin, $\chi_{\rm eff}$, and effective precession spin, $\chi_{\rm p}$, for the local AGN BBH-merger population. Compared to the fiducial spin models presented in the main text, the most striking difference is the disappearance of the pronounced bimodality in $\chi_{\rm eff}$ for mixed prograde-retrograde populations ($f_{\rm progr}=0.5$). In the fiducial case, this bimodality originates from the preferential alignment or anti-alignment of BH spins with the disk angular momentum, combined with non-negligible natal spin magnitudes. When the initial spins are strongly suppressed, the contribution from first-generation mergers clusters around $\chi_{\rm eff}\simeq0$, eliminating the double-peak structure.

The sign of $\chi_{\rm eff}$ still retains information about the underlying orbital orientation distribution at BBH formation. Populations with $f_{\rm progr}=1$ preferentially populate positive $\chi_{\rm eff}$ values, while populations with $f_{\rm progr}=0$ preferentially populate negative values. The mixed case, $f_{\rm progr}=0.5$, remains symmetric around $\chi_{\rm eff}=0$. 

The distributions of $\chi_{\rm p}$ are comparatively less affected by the change in the natal-spin prescription. In all cases, $\chi_{\rm p}$ displays a sharp peak at zero and a tail extending up to $\chi_{\rm p}\simeq 0.8$. 

Overall, these results show that the detailed morphology of the spin distributions, particularly the presence or absence of bimodality in $\chi_{\rm eff}$, depends sensitively on the natal spins of stellar-origin BHs.

\section{Gas capture}\label{app:gas_capture}
The BHs are initially assumed to orbit in the NSC with isotropically distributed inclinations with respect to the AGN disk plane. Gas drag is expected to damp both the inclination $i$ and the eccentricity $e$ of their orbit \citep{Cresswell_2007}. We therefore sample the initial inclination isotropically and define the dimensionless inclination as $\tilde{i} = \sin{i}/h$, where $h=H/R$ is the local disk aspect ratio.

For BHs initially orbiting outside the AGN disk plane, corresponding to $\tilde{i}>1$, we model the capture process using the fitting formulae derived by \citet{Rowan_2025} from dedicated hydrodynamical simulations. 
The inclination damping efficiency is \citep[][their eq. 46]{Rowan_2025} as
\begin{equation}
\Phi (\tilde{i}) = \frac{\Delta \tilde{i}}{\tilde{i}} =
\begin{cases}
A & \tilde{i} < \tilde{i}_{\rm c}, \\
B\,\tilde{i}^{-2.64} & \tilde{i} \geq \tilde{i}_{\rm c},
\end{cases}
\end{equation}
where $\tilde{i}_{\rm c}=4.6$, $\log_{10}A = 0.67\log_{10}\tilde{m}_{\rm H} - 2.64\log_{10}\tilde{i}_{\rm c} + 1.80$, and $\log_{10}B = 0.67\log_{10}\tilde{m}_{\rm H} + 1.80$. The quantity $\tilde{m}_{\rm H}$ is the dimensionless ambient Hill mass, $\tilde{m}_{\rm H} = \pi R_{\rm H}^2 \Sigma_{\rm g} / m_1$, where $R_{\rm H} = R (m_1/3\MSMBH)^{1/3}$ is the Hill radius of the BH.

Assuming that the BH crosses the disk twice in its period, $T = 2 \pi / \Omega_\mathrm{K}$, the characteristic damping timescale is then 
\begin{equation}
    t_{\rm damp} = \frac{\tilde{i}}{|\Delta \tilde{i}/\Delta t|} = \frac{\pi}{\Omega_\mathrm{K}}\, \frac{1}{\Phi(\tilde{i})},
\end{equation}
where $\Delta t=T/2$ and $\Omega_\mathrm{K} = \sqrt{G\MSMBH/R^3}$ is the local Keplerian frequency. 

The fitting formulae from \citet{Rowan_2025} are calibrated on initially circular orbits. \citet{Wang_2024} show that the orbital eccentricity is generally damped on a timescale comparable to, or shorter than, the inclination damping timescale, such that the orbit is approximately circular by the time disk capture becomes efficient. 

\section{Binary pairing}\label{app:pairing}
We allow for pair-ups between BHs of arbitrary generations, denoted as $N$g--$M$g mergers. Following \citet{Zevin-Holz}, we assume that the probability of selecting a secondary of generation $M$ scales as $p(M)\propto 2^{-(M-1)}$, with $M\leq N$. If the BBH merges, the generation of the remnant is defined as $N^\prime = N+M$.
If $M=1$, we sample the BH mass as in \citet{fastcluster2021}, $p(m_2|m_1)\propto (m_1+m_2)^4$, between $5\Msun$ and $m_1$. Otherwise, we recursively build the BH through a sequence of mergers, computing the remnant mass and spin at each step following \citet{Jimenez}. We set the secondary BH spin magnitude and tilt in the same way as for the primary BH.
Following the results of few-body simulations \citep[][their eq.s~7, 12]{Qian_2024}, we take the pairing timescale to be
\begin{equation}
t_{\rm pair} = \frac{3\, c_{\rm s}^3\, h\, R}{4\, \pi\, G^2\, \Sigma_{\rm g}\, m_1},
\label{eq:t_pair}
\end{equation}
and assume that BBH formation occurs only when
$t_{\rm pair} \leq 5.3\, t_{\rm K}$,
where $t_{\rm K}=2\pi/\Omega_{\rm K}$ is the local Keplerian period. We note that this condition provides only a necessary requirement for binary formation, but not a sufficient one, as a bound BBH is not guaranteed to form even when the timescale criterion is satisfied \citep{Qian_2024}. We sample the initial semimajor axis from $p(a)\propto a^{9/2}$ between $1\,R_\odot$ and $R_\mathrm{Hill}/2$ \citep{Binney-Tremaine-2008}, and eccentricities from a thermal distribution $p(e)=2e$ for $e$ between 0 and 1 \citep{Jeans_1919_binaries}.

The geometry of BBH formation inside AGN disks remains uncertain, as both few-body and hydrodynamical simulations suggest that binary formation may depend sensitively on the orbital configuration of the interacting BHs and on the local properties of the gas \citep{Qian_2024, Whitehead_2024}. To capture these uncertainties in a simple way, we introduce a parameter $f_{\rm progr}$, representing the fraction of BBHs whose internal orbital angular momentum is aligned with the angular momentum of the surrounding AGN disk at formation.
We assume that $f_{\rm progr}$ is constant throughout the disk and treat it as a free parameter of the model. This allows us to explore limiting cases ranging from fully prograde ($f_{\rm progr}=1$) to mixed ($f_{\rm progr}=0.5$) or fully retrograde populations ($f_{\rm progr}=0$).

\section{Binary orbital evolution}
\label{app:binary_evolution}

%%%%%%%%%%%%%%%%%%% FIGURE %%%%%%%%%%%%%%%%%%%%%
\begin{figure}
    \centering
    \includegraphics[width=\linewidth]{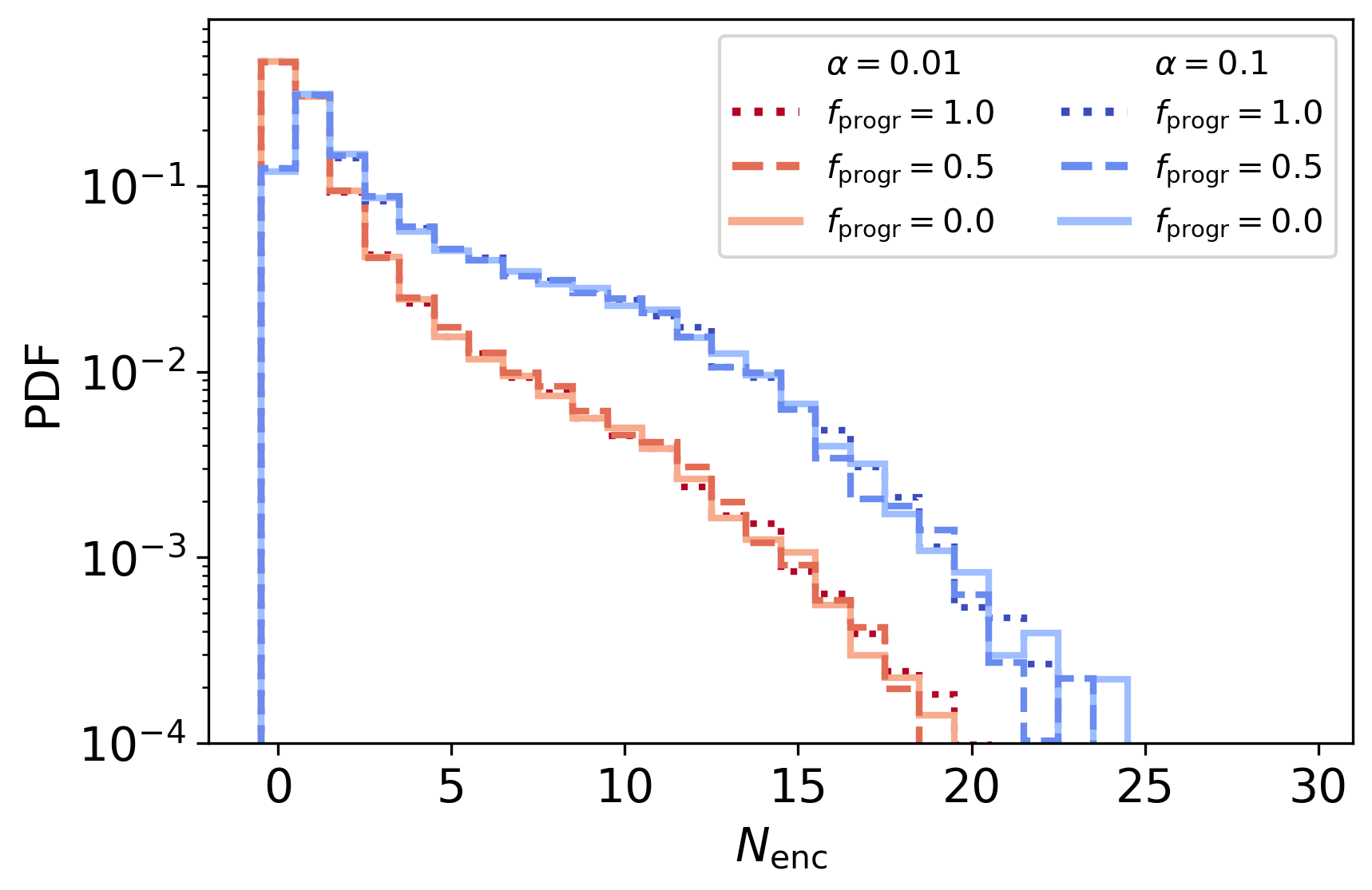}
    \caption{Distribution of the number of binary--single encounters, $N_{\rm enc}$, experienced by merging BBHs in the $\sigma_{\rm BH}\ll1$ case. Histograms show the probability density function of the local populations for the two disk viscosities, $\alpha=0.01$ (orange) and $\alpha=0.1$ (blue), and for different assumptions on the fraction of binaries born prograde with respect to the AGN disk: $f_{\rm progr}=1.0$ (dotted), $0.5$ (dashed), and $0.0$ (solid). }
    \label{fig:Nencounter}
\end{figure}
%%%%%%%%%%%%%%%%%%%%%%%%%%%%%%%%%%%%%%%%%%%%%%%%%

\subsection{Gas-driven evolution}

In the framework of \citet{Ishibashi_2024}, a binary of total mass $m_{\rm b}=m_1+m_2$ and mass ratio $q=m_2/m_1$ evolves its semi-major axis, $a$, and eccentricity, $e$, through the combined effect of gas torques and accretion. The gas-driven evolution is given by
\begin{align}
\frac{\dot{a}_{\rm g}}{a} =\;&
\frac{2l}{m}
\frac{\Gamma_{\rm acc}-\Gamma_{\rm visc}}
{\mu a^2 \Omega_{\rm b}}
+
\frac{\dot{m}_{\rm b}}{m_{\rm b}}
\left[
1+f(q)
\right],
\\
\frac{\dot{e}_{\rm g}\, e}{1-e^2} =\;&
\frac{\Gamma_{\rm visc}-\Gamma_{\rm acc}}
{\mu a^2 \Omega_{\rm b}}
\left(
\frac{1}{\sqrt{1-e^2}}-\frac{l}{m}
\right)
+
\frac{\dot{m}_{\rm b}}{m_{\rm b}}
\left[
1+\frac{3\, f(q)}{2}
\right],
\end{align}
where $\mu = m_1 m_2/m_{\rm b}$ is the binary reduced mass, $\Omega_{\rm b}$ is the binary orbital frequency, and we adopt $(m,l)=(2,1)$, corresponding to the dominant Lindblad resonance.

The viscous torque exerted by the circumbinary disk is
$\Gamma_{\rm visc} =
4 a^2 (1+e)^2 \Omega_{\rm K} \dot{m}_{\rm b}$,
while the accretion torque is
\begin{equation}
\Gamma_{\rm acc} =
\frac{2\pi}{\Omega_{\rm b}}
\frac{\dot{m}_{\rm b}}{1+q_a}
\left[
\frac{G m_{\rm b}}{1+q}
\frac{\alpha h^2}{0.3 a (1-e)}
\left(
q^{0.3}+q^{0.7}q_a
\right)
\right],
\end{equation}
in the assumption that both the circumbinary disk and the mini-disks surrounding the individual BBH components have the same viscosity parameter, $\alpha$, and aspect ratio, $h$, as the ambient AGN disk at the binary location.

The mass inflow rate through the circumbinary disk is taken as
\begin{equation}
\dot{m}_{\rm b}
=
\frac{4\pi G h}
{c\kappa}
\sqrt{
\MSMBH m_{\rm b}
\frac{2a(1+e)}{R}
} ,
\end{equation}
where $R$ is the orbital radius of the BBH within the AGN disk, $\dot{m}$ is the dimensionless accretion rate, and $\kappa=0.34\,{\rm cm^2\,g^{-1}}$ is the electron-scattering opacity.
Finally, we define
\begin{equation}
f(q) = \frac{q^2+q_a}{q(1+q_a)}, \quad q_a = \frac{1}{0.1+0.9q}. 
\end{equation}

The binary also hardens due to the effect of GW  emission, which will govern the evolution at small semimajor axes. The evolution of the semimajor axis, $\dot{a}_\mathrm{GW}$, and eccentricity, $\dot{e}_\mathrm{GW}$, due to GW hardening proceeds as in \citet{Peters_1964}. The overall evolution of the binary is thus described by eq.~\ref{eq:IG and Peters}.

In practice, we evolve the BBH semi-major axis and eccentricity under gas hardening until GW emission becomes dominant. At each timestep, we compute the hardening and eccentricity-growth rates from the circumbinary disk model and combine them with the standard GW-driven evolution equations. The inspiral is followed until either the BBH merges or the total delay time exceeds the AGN lifetime.

\subsection{Binary-single encounters}

We estimate the characteristic encounter timescale as \citep{Leigh_McKernan_2018}
\begin{equation}
    t_\mathrm{enc} =  \frac{\MSMBH^{3/2}\,  h^2(R_\mathrm{max})}{m_*\, N_*\, \Sigma_\mathrm{g}(R_\mathrm{max}) \, R_\mathrm{max}^{1/2} },
    \label{eq:t_enc}
\end{equation}
where $m_* = 1 \Msun$ and $N_*$ is the number of objects embedded in the disk, estimated as in \citet[][eq. 7]{Vaccaro_2023}.
We compare the encounter timescale to the gas-driven inspiral timescale, to determine whether the binary undergoes any encounter before merger. 

To model these interactions, we resort to a pre-computed grid of three-body scattering experiments performed with the post-Newtonian few-body code \tsunami{} \citep{tsunami}. The grid spans different SMBH masses, BBHs from first to highly hierarchical generations ($1g$--$500g$), both prograde and retrograde configurations, and two velocity dispersion regimes (dynamically cold or dynamically hot). 

The disc velocity dispersion is parametrized with a single dimensionless parameter, $\sigma_{\rm BH}=\sigma_{e}=2\sigma_{i}$, where $\sigma_{e}$ and $\sigma_{i}$ are the scale parameters of the Rayleigh distributions of eccentricity and inclination. In this study, we consider the two extremal regimes shown in \citet{Trani_2024}. We adopt as fiducial model a dynamically cold regime, corresponding to $\sigma_{\rm BH} < 10^{-4}$, where objects' orbits are already strongly damped, deeply embedded in gas, and nearly coplanar. We also consider in our discussion section a dynamically warmer regime, $\sigma_{\rm BH}=0.1$, which may arise when stochastic perturbations, such as turbulence, compete with gas damping and maintain finite eccentricities and inclinations.

Inner binary separations are sampled from $p(a)\propto a^{9/2}$ between $1\,R_\odot$ and $R_\mathrm{Hill}/2$, eccentricities from a thermal distribution $p(e)=2e$, and all encounters are initialized at a distance of $10^4\,R_\mathrm{S}$ from the SMBH.
For each encounter, we read the outcome from the \tsunami{} grid and update the BBH orbital properties accordingly.  

%The possible outcomes include: shrinking or widening the binary, increasing its eccentricity, tilting its orbital plane (prograde-to-retrograde or retrograde-to-prograde flips), exchanging one of its components, ionizations, or direct collisions during the encounter.
If the encounter ionizes the binary, we stop the evolution and classify the BBH as disrupted. If the encounter leads to a direct collision, we classify the system as a plunge-in event. If an exchange occurs, we replace the secondary BH with a radomly-drawn $1g$ BH and continue the evolution with the updated BBH masses and spins.
For surviving BBHs, after each encounter we reset the binary semi-major axis, $a$ and eccentricity, $e$ to the post-encounter values. We then re-integrate the inspiral using the same gas-hardening and GW prescriptions as in eq.~\ref{eq:IG and Peters}. If the BBH reaches the merger condition, the inspiral time is updated to eq.~\ref{eq:inspiral_timescale}. The distribution of the number of encounters a BBH goes through before merger, for astrophysical populations constructed as in \autoref{sec:cosmo_methods}, is shown in \autoref{fig:Nencounter}. Most binaries undergo only a small number of interactions, with the distribution peaking at 0 for $\alpha=0.01$ and 1 for $\alpha=0.1$. However, repeated encounters are possible, as the distribution presents a tail extending to $N_{\rm enc}\sim 20$--$25$.

\section{Outcomes of binary-single encounters}\label{app:3bb}
%%%%%%%%%%%%%%%%%%%%%% FIGURES 3BB %%%%%%%%%%%%%%%%%%%%%%%%%%%%%%

\begin{figure}
    \centering
    \includegraphics[width=\linewidth]{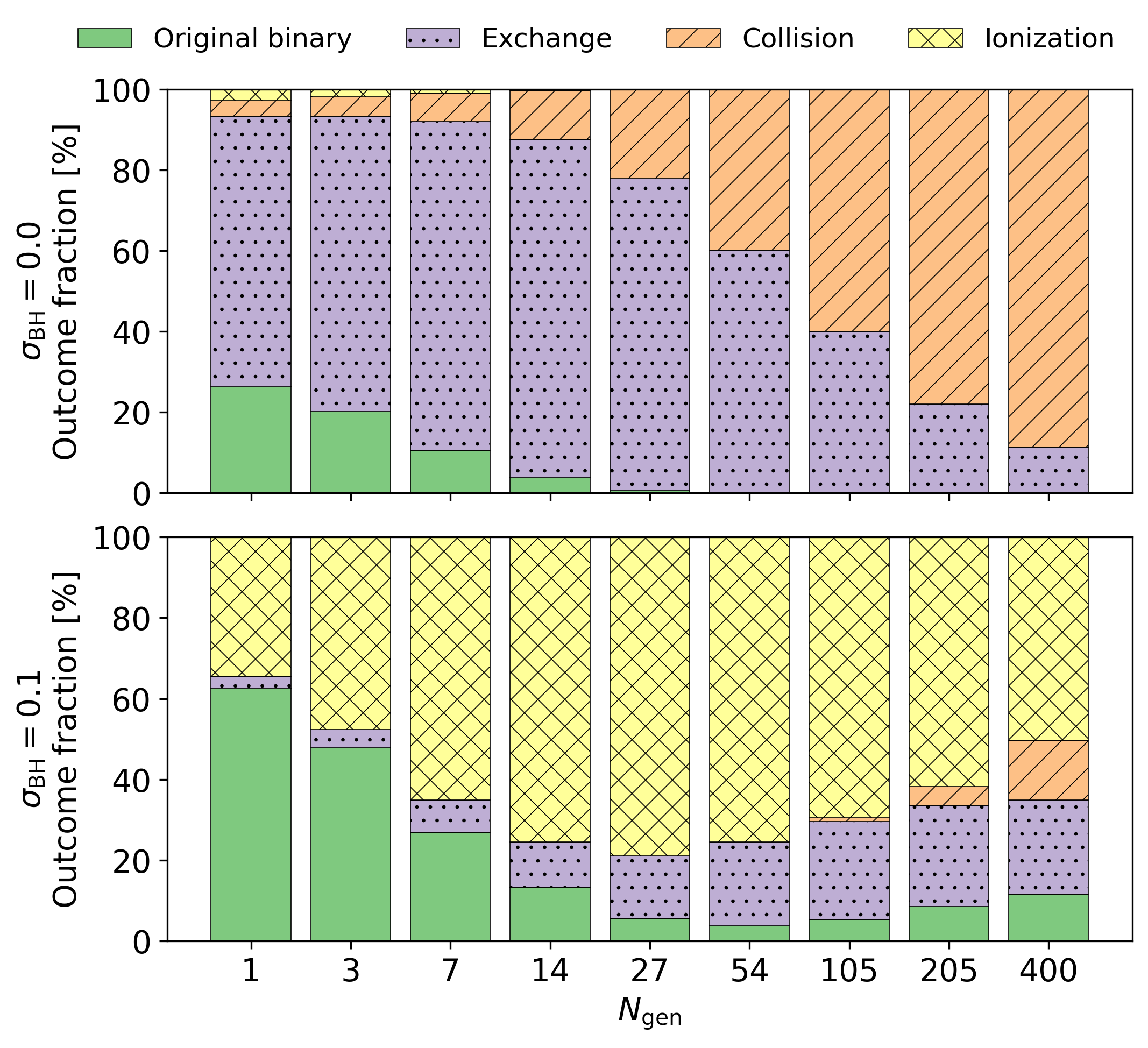}
    \caption{Outcome fractions of binary--single encounters as a function of primary generation number, $N_{\rm gen}$, for the cold-disk ($\sigma_{\rm BH}=0.0$, top) and hot-disk ($\sigma_{\rm BH}=0.1$, bottom) scenarios. Each stacked bar shows the relative contribution of the possible encounter outcomes: preservation of the original binary, exchange, collision, and ionization.}
    \label{fig:encounter_outcome_stacked_bars}
\end{figure}

\begin{figure*}
    \centering
    \includegraphics[width=\linewidth]{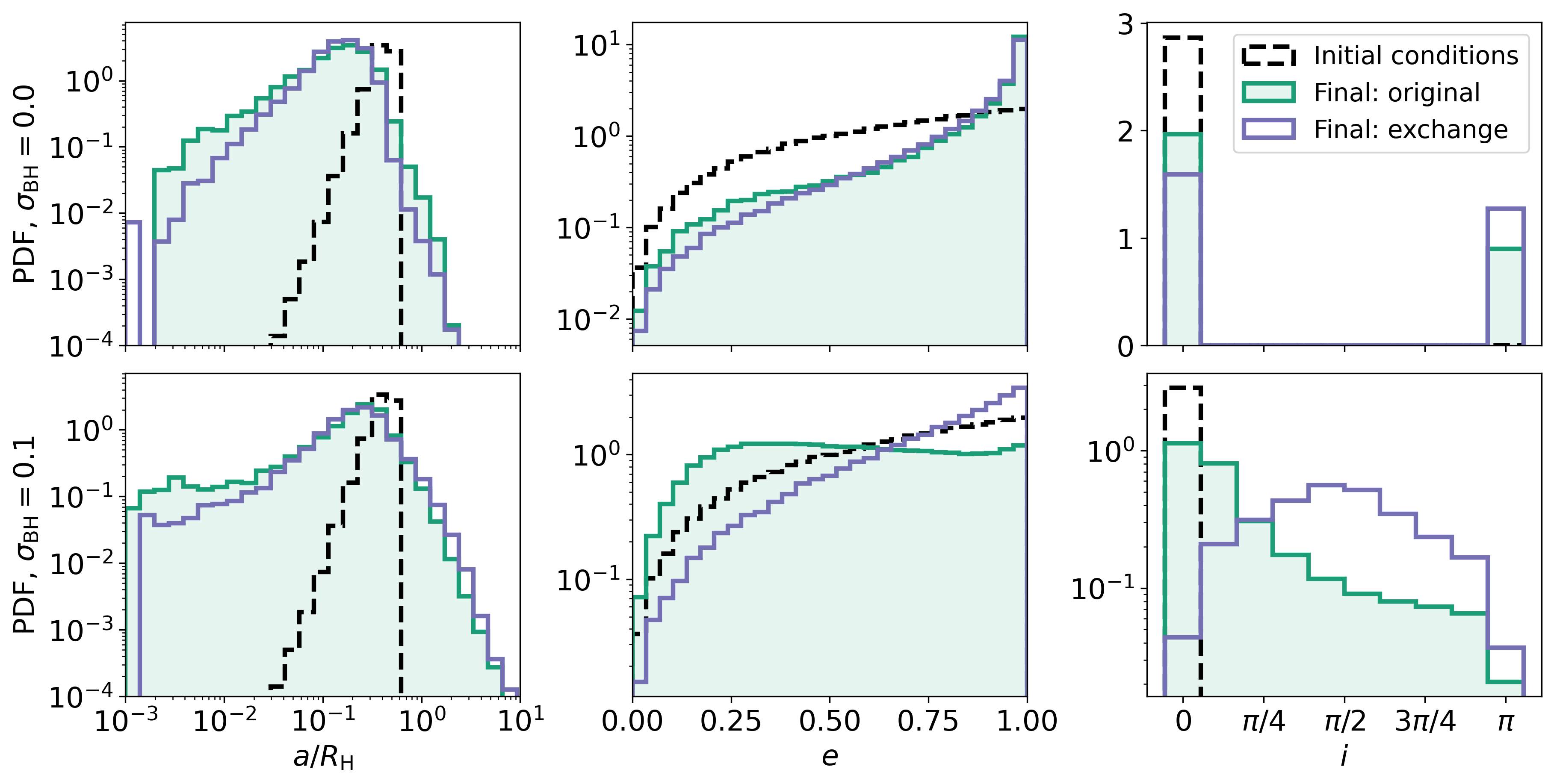}
    \caption{Initial and final distributions of BBH orbital properties for surviving binaries after binary--single encounters in the cold-disk ($\sigma_{\rm BH}=0.0$, top row) and hot-disk ($\sigma_{\rm BH}=0.1$, bottom row) scenarios. Columns show the semi-major axis in units of Hill radius, $a/R_{\rm H}$, eccentricity, $e$, and inclination, $i$. Dashed black histograms indicate the initial conditions, while solid green and purple histograms show the final distributions of binaries that preserve the original pairing and binaries produced through exchange, respectively.}
    \label{fig:encounter_initial_final_histograms_Hill}
\end{figure*}
%%%%%%%%%%%%%%%%%%%%%%%%%%%%%%%%%%%%%%%%%%%%%%%%%%%%%%%%%%%%%%%%%%%

In this Appendix, we summarize the outcomes of binary--single interactions in our AGN-disk population model. 
The underlying encounter outcomes are based on the scattering experiments of \citet{Trani_2024}, but are evaluated using the initial conditions relevant for our AGN-disk population synthesis as detailed in \autoref{app:binary_evolution}.

The distributions of post-encounter outcomes for the cold and warm disk models, shown for different generations of the primary BH, are shown in \autoref{fig:encounter_outcome_stacked_bars}.
We classify the outcome of each binary--single encounter into four mutually exclusive categories: 
\begin{enumerate}[(i)]
    \item a surviving original binary, in which the two initial binary members remain bound to each other; 
    \item an exchange, in which the binary survives but one of its original members is replaced by the incoming single BH; 
    \item a collision, corresponding to a close physical encounter between two compact objects during the three-body interaction;
    \item an ionization, in which the original binary is distrupted and no bound BBH remains.
\end{enumerate}

In the present model, collisions are not counted as BBH mergers: our merger sample includes only binaries whose coalescence is obtained by explicitly evolving the BBH inspiral under gas hardening and GW emission. A collision recorded in the binary--single scattering calculation, instead, indicates that two compact objects undergo a close encounter during the three-body interaction itself, but the subsequent evolution of such an event is not resolved within our semi-analytical framework. For this reason, we do not assign a merger time to collisional outcomes and exclude them from the merger population. Any prompt mergers associated with these encounters would therefore constitute an additional contribution not included in the populations reported here.

For systems that remain bound after the interaction, the encounter can either harden or soften the binary, as shown in \autoref{fig:encounter_initial_final_histograms_Hill}. 
In most cases, the post-encounter separation is smaller than the pre-encounter value, indicating that binary--single interactions often promote subsequent gas- or GW-driven inspiral. However, a subset of systems is softened instead, emerging from the encounter with a larger semi-major axis. 
In our model, softened systems are retained only if their final separation remains below the survival threshold, $a<R_{\rm H}/2$, while binaries expanded beyond this limit are treated as disrupted, since their mutual binding is too weak compared to the tidal field of the SMBH.

The eccentricity distribution of surviving binaries is generally excited by binary--single interactions, with the exception of original binaries in the $\sigma_{\rm BH}=0.1$ case.
The post-encounter inclination distribution shows that encounters can tilt binaries away from their original orientation. Specifically, in the coplanar limit ($\sigma_{\rm BH}=0.0$), the interaction preferentially flips the binary orientation, switching between prograde and retrograde configurations.  
For $\sigma_{\rm BH}=0.1$, instead, original binaries instead develop a broader inclination distribution while still retaining partial memory of their initial alignment. 
Exchanged binaries are much rarer in this case, but their orientations are more nearly isotropic, reflecting the fact that the post-encounter binary is effectively reassembled during the interaction.

\section{Features in the mass-ratio distribution}\label{app:q_Ngen}

\begin{figure}
    \centering
    \includegraphics[width=0.87\linewidth]{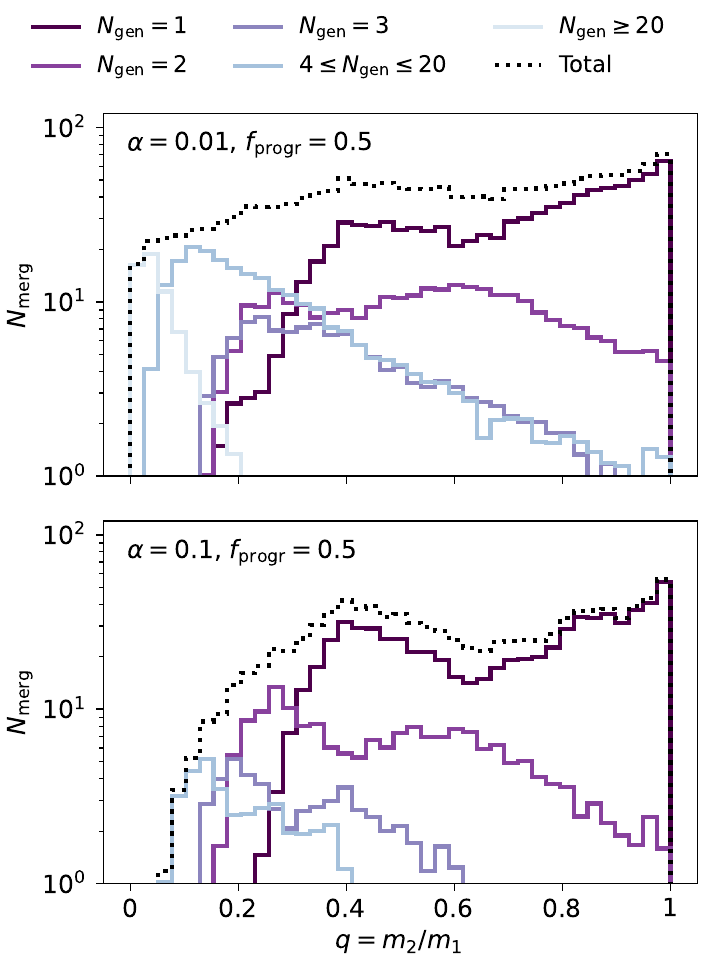}
    \caption{Mass-ratio distributions of mergers in the local weighted AGN population, separated by the generation number of the primary BH. The top and bottom panels show models with $\alpha=0.01$ and $\alpha=0.1$, respectively, for $f_{\rm progr}=0.5$. 
Solid curves show the contribution from different hierarchical-merger generations, while the dotted black curve shows the total distribution. }
    \label{fig:q_study}
\end{figure}

\autoref{fig:q_study} shows that the total mass-ratio distribution has two main features: a peak near equal masses, $q\simeq1$, and a broader enhancement at $q\simeq0.3$--$0.7$, most visible for $\alpha=0.1$. 
When the population is separated by the generation number of the primary BH, both features are mainly associated with first-generation binaries. 
They therefore reflect the adopted initial stellar-origin BH mass spectrum and companion-assignment prescription, rather than hierarchical growth alone.

In our simulations, first-generation BH masses are drawn from \sevn{} population-synthesis models at $Z=0.02$ \citep{sevn_2023}. 
The corresponding mass spectrum contains preferred mass ranges, including a broad peak at $m_1\simeq7$--$10\,\Msun$ and sharper peaks around $m_1\simeq19\,\Msun$ and $m_1\simeq22\,\Msun$. 
Pairing BHs drawn from similar mass ranges naturally contributes to the peak at $q\simeq1$, while pairing BHs from different characteristic mass ranges can populate the secondary feature at lower $q$.

At larger $N_{\rm gen}$, the mass-ratio distribution becomes progressively skewed toward lower values of $q$. 
This reflects the hierarchical growth of the primary BH, while the secondary is more often drawn from a lower-generation population. 
When hierarchical mergers are efficient, as in the $\alpha=0.01$ models, this low-$q$ extension can partially mask the first-generation feature around $q\simeq0.4$.

\section{Timescales}\label{app:timescales}

\begin{figure*}
    \centering
    \includegraphics[width=\linewidth]{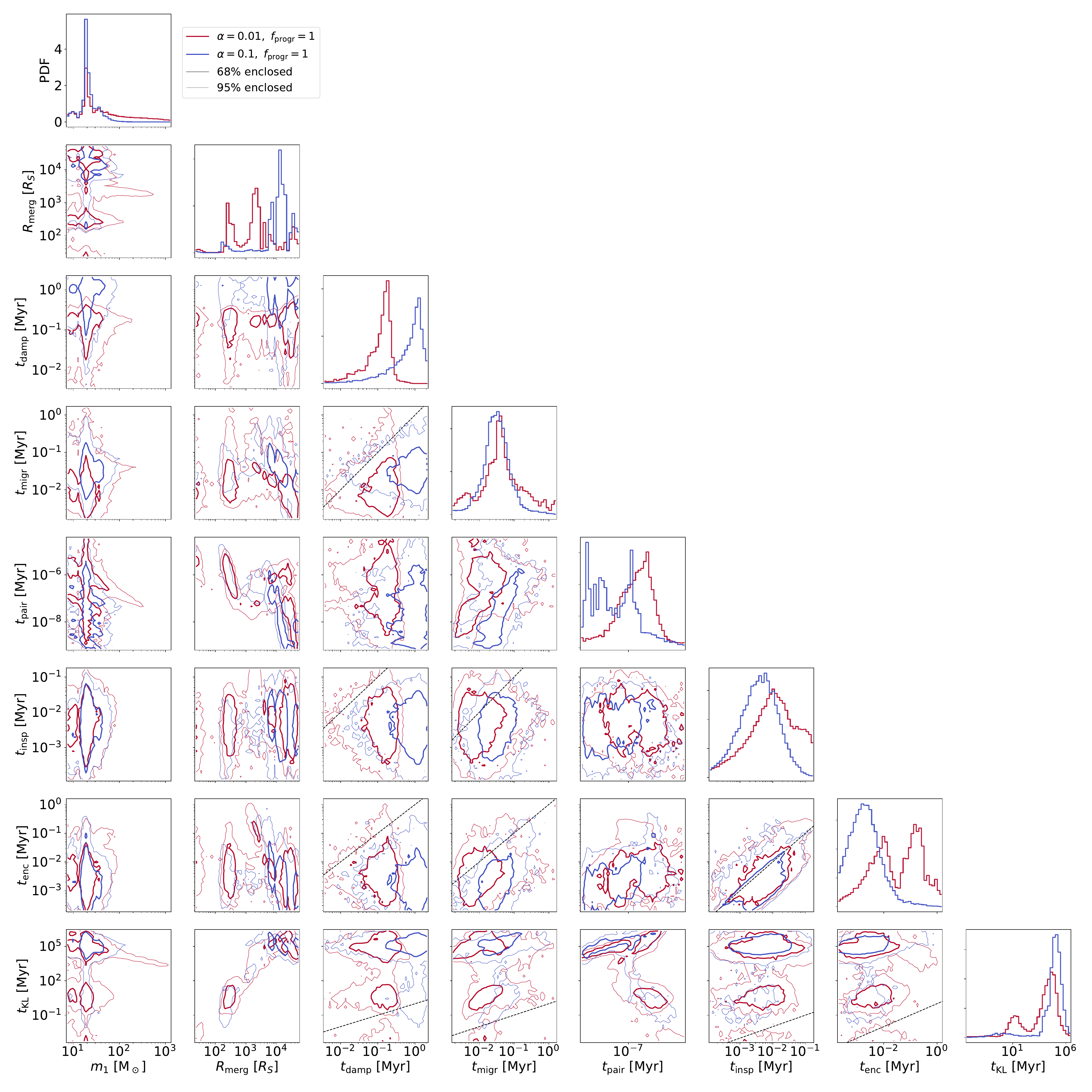}
    \caption{Corner plot of the main formation and evolutionary timescales for the intrinsic local AGN population, shown for the fully prograde case ($f_{\rm progr}=1$). Red and blue contours correspond to $\alpha=0.1$ and $\alpha=0.01$, respectively.  Diagonal panels show the one-dimensional probability density functions, while off-diagonal panels show the corresponding two-dimensional distributions. Thick and thin contours enclose $68\%$ and $95\%$ of the probability density, respectively. The quantities shown are the primary mass $m_1$, merger radius $R_{\rm merg}$, capture time $t_{\rm damp}$, migration time $t_{\rm migr}$, pairing time $t_{\rm pair}$, inspiral time $t_{\rm insp}$, encounter time $t_{\rm enc}$, and timescales of von~Zeipel-Kozai-Lidov oscillations, $t_{\rm KL}$. }
    \label{fig:corner_contours_timescales}
\end{figure*}

\autoref{fig:corner_contours_timescales} compares the main timescales entering the BBH evolution in our AGN disk models for the local astrophysical population. 
The damping time, $t_{\rm damp}$, and the migration time, $t_{\rm migr}$, span several orders of magnitude, reflecting the fact that the local population is assembled by weighting simulations over a wide range of SMBH masses and disk locations. $t_{\rm damp}$ peaks at roughly $0.1$ Myr ($1$ Myr) in the case with $\alpha = 0.01$ ($0.1$), whereas $t_{\rm migr}$ peaks at roughly $4\times 10^{-2}$ Myr.

The pair-up timescale, $t_{\rm pair}$, is generally short compared to the other relevant timescales, with typical values in the range $10^{-9}$--$10^{-5}$ Myr. This suggests that, once two BHs are brought sufficiently close by damping and migration, gas-assisted pairing is not the bottleneck of the evolution. Lower values of the pair-up timescales ($t_{\rm pair}\lesssim10^{-7}$ Myr) are associated with mergers occurring in the outer disk ($R_{\rm merg}\gtrsim10^4 R_{\rm S}$), where the pair-up condition $t_{\rm pair}\lesssim5.3 t_{\rm K}$ \citep{Qian_2024} is more easily satisfied, as the local Keplerian period increases with more sharply with radius than the pairing timescale.

The inspiral time, $t_{\rm insp}$, controls the final hardening of the BBH after formation. 
It typically lies in the range $10^{-4}$--$10^{-1}$ Myr, with the distribution peaking at $t_{\rm insp}\simeq 10^{-2}$ ($3\times10^{-3}$) Myr in the $\alpha=0.01$ ($0.1$) case.
The encounter time, $t_{\rm enc}$, spans a comparable range of valuesand peaks at $t_{\rm enc}\simeq 10^{-1}$ ($3\times10^{-3}$) Myr in the $\alpha=0.01$ ($0.1$) case.
It is interesting to notice that, although the order of magnitudes of the inspiral and encounter timescale are comparable, the bulk of the population falls in the area where $t_{\rm insp} < t_{\rm enc}$. This implies that, in most cases, gas-driven hardening is fast enough for the BBH to merge before a binary--single encounter occurs (see also \autoref{app:binary_evolution}).

For completeness, we also estimate the von~Zeipel-Kozai-Lidov timescale associated with the hierarchical triple formed by the BBH and the central SMBH \citep{Kozai_1962, Lidov_1962}. 
We treat the BBH as the inner binary, with total mass $m=m_1+m_2$,
semi-major axis $a$, and orbital period $P_{\rm in}= 2\pi\sqrt{a^3/Gm}$.
The outer orbit is the motion of the BBH centre of mass around the SMBH, with semi-major axis approximated by the radial location $R$, and period $P_{\rm out}= 2\pi\sqrt{R^3/G\MSMBH}$ (assuming $m\ll \MSMBH$).
At quadrupole order, the characteristic von~Zeipel-Kozai-Lidov time is estimated as \citep{Naoz_2016}
\begin{equation}
    t_{\rm KL}
    =
    \frac{8}{15\pi}
    \left( \frac{\MSMBH+m}{\MSMBH} \right)
    \frac{P_{\rm out}^2}{P_{\rm in}}
    \left(1-e_{\rm out}^2\right)^{3/2} \simeq \frac{8\,P_{\rm out}^2}{15\pi\,P_{\rm in}},
    \label{eq:t_KL}
\end{equation}
assuming $m\ll \MSMBH$ and $e_{\rm out}=0$.

As shown in \autoref{fig:timescales_for_discussion}, $t_{\rm KL}$ is typically longer than other relevant timescales, including the the gas-driven and encounter timescales relevant for BBH evolution. von~Zeipel-Kozai-Lidov oscillations would not have enough time to significantly modify the binary eccentricity or inclination before the binary is damped, hardened, perturbed by encounters, or merged, with the sole exception of BBHs merging in the innermost region of the disk, at $R_{\rm merg}\lesssim10 R_{\rm S}$. Therefore, we do not include von~Zeipel-Kozai-Lidov evolution explicitly in the fiducial model.

\end{document}